\documentclass[twocolumn,resetfootnote]{aastex701}

\begin{document}
\title{CHANG-ES. {XXXVIII}. A Thin Radio Halo Shaped by Slow Cosmic-Ray Transport in the Quiescent Galaxy NGC~4565}

\author[0009-0006-3887-8988, gname=Jianghui, sname=Xu]{Jianghui Xu}
\affiliation{Department of Astronomy, University of Science and Technology of China, Hefei, Anhui 230026, People's Republic of China}
\affiliation{School of Astronomy and Space Science, University of Science and Technology of China, Hefei 230026, People's Republic of China}
\email{jianghuixu@mail.ustc.edu.cn}

\author[0000-0001-6239-3821, gname=Jiang-Tao, sname=Li]{Jiang-Tao Li}
\affiliation{Purple Mountain Observatory, Chinese Academy of Sciences, 10 Yuanhua Road, Nanjing 210023, China}
\email[show]{pandataotao@gmail.com}

\author[0000-0003-4286-5187, gname=Guilin, sname=Liu]{Guilin Liu}
\affiliation{Department of Astronomy, University of Science and Technology of China, Hefei, Anhui 230026, People's Republic of China}
\affiliation{School of Astronomy and Space Science, University of Science and Technology of China, Hefei 230026, People's Republic of China}
\email[show]{glliu@ustc.edu.cn}

\author[0009-0008-2940-6166, gname=Luan, sname=Luan]{Luan Luan}
\affiliation{Purple Mountain Observatory, Chinese Academy of Sciences, 10 Yuanhua Road, Nanjing 210023, China}
\email{luanluan@pmo.ac.cn}

\author[0000-0002-2082-407X, gname=Volker, sname=Heesen]{Volker Heesen}
\affiliation{Hamburger Sternwarte, Universit\"at Hamburg, Gojenbergsweg 112, D-21029 Hamburg, Germany}
\email{volker.heesen@uni-hamburg.de}

\author[0009-0001-8154-3562, gname=Rainer, sname=Beck]{Rainer Beck}
\affiliation{Max-Planck-Institut f\"ur Radioastronomie, Auf dem H\"ugel 69, 53121 Bonn, Germany}
\email{rbeck@mpifr-bonn.mpg.de}

\author[0000-0003-0073-0903, gname=Judith, sname=Irwin]{Judith Irwin}
\affiliation{Department of Physics, Engineering Physics, \& Astronomy, Queen's University, Kingston, Ontario K7L 3N6, Canada}
\email{irwinja@queensu.ca}

\author[0000-0002-9279-4041, gname='Q. Daniel', sname=Wang]{Q. Daniel Wang}
\affiliation{Department of Astronomy, University of Massachusetts, Amherst, MA 01003, USA}
\email{wqd@umass.edu}

\author[0000-0001-8428-7085, gname=Michael, sname=Stein]{Michael Stein}
\affiliation{Ruhr University Bochum, Faculty of Physics and Astronomy, Astronomical Institute (AIRUB), 44780 Bochum, Germany}
\email{mstein@astro.ruhr-uni-bochum.de}

\author[0000-0002-3286-5346, gname=Li-Yuan, sname=Lu]{Li-Yuan Lu}
\affiliation{Department of Physics and Astronomy \& Research Center of Astronomy, Qinghai University, 251 Ningda Road, Xining 810016, People’s Republic of China}
\email{lylu@qhu.edu.cn}

\author[0000-0001-7254-219X, gname=Yang, sname=Yang]{Yang Yang}
\affiliation{Xiangtan University, Xiangtan 411105, Hunan, People’s Republic of China}
\email{yangyang.astro@gmail.com}

\author[0000-0003-2623-2064, gname=Jeroen, sname=Stil]{Jeroen Stil}
\affiliation{Department of Physics and Astronomy, the University of Calgary, 2500 University Drive NW, Calgary, AB T2N 1N4, Canada}
\email{jstil@ucalgary.ca}

\author[0000-0001-5310-1022, gname=Jayanne, sname=English]{Jayanne English}
\affiliation{Department of Physics \& Astronomy, University of Manitoba, Winnipeg, Manitoba, R3T 2N2, Canada}
\email{jayanne.english@umanitoba.ca}

\author[0000-0001-8206-5956, gname=Ralf-J\"urgen, sname=Dettmar]{Ralf-J\"urgen Dettmar}
\affiliation
{Ruhr University Bochum, Faculty of Physics and Astronomy, Astronomical Institute (AIRUB), 44780 Bochum, Germany}
\email{dettmar@astro.ruhr-uni-bochum.de}

\correspondingauthor{Jiang-Tao Li, Guilin Liu}

\begin{abstract}
We present the VLA C-array $S$-band (2--4 GHz) radio continuum observations of the nearby edge-on spiral galaxy NGC~4565, a target from the Continuum Halos in Nearby Galaxies—an EVLA (CHANG-ES) Survey. 
We conduct rotation measure synthesis to probe the magnetic field structure and analyze the vertical radio continuum intensity profiles using the 1-D cosmic ray transportation models.
The radio continuum emission of NGC~4565 is vertically compact, with a vertical-to-radial extent ratio of $\sim 1/6$. Its vertical profile is optimally described by a two-component Gaussian distribution, yielding a mean Gaussian halo scale height of $\sim 3.0\,{\rm kpc}$.
The magnetic field is weak, predominantly disk-parallel, with an equipartition strength of $\lesssim5\,\mu$G and a rotation measure profile indicative of an axisymmetric spiral structure.
Nevertheless, we identify a localized, faint vertical magnetic field component in the northeastern region, hinting at an X-shaped structure that spatially coincides with extraplanar structures detected in \ion{H}{1} and soft X-ray emission.
The CR transport modeling favors a flux-tube advection scenario, with a slow initial velocity of $v_0 \approx 60\,{\rm km\,s^{-1}}$, consistent with a limited energy input from star formation. 
Therefore, the absence of an extended radio halo can be explained by the low star formation rate, the weak magnetic field, and the inefficient CR transport. The localized X-shaped field may trace a weak, magnetically guided outflow or a tidal perturbation induced by the nearby companion. 
NGC~4565 is thus a key quiescent benchmark for understanding the physical conditions required to drive large-scale outflows and generate extended radio halos.
\end{abstract}


\section{Introduction}
\label{sec:introduction}

Magnetic fields and cosmic rays (CRs) are important components of the interstellar and circumgalactic medium (ISM and CGM), and they play a marked role in shaping the physical conditions of galaxies. Magnetic fields provide additional pressure support against gravity from the galaxy \citep{Boulares90}, regulate the transport of CRs \citep{Becker20}, and can guide the gas flow from the galactic disk to the halo \citep{Girichidis18, Veilleux20}. Cosmic rays, accelerated and injected by powerful astrophysical sources such as supernova remnants and active galactic nuclei (AGN), are closely coupled to the magnetic fields and propagate throughout the ISM, transferring momentum and energy to the halo. These two closely associated components are crucial for maintaining the dynamical balance of the ISM \citep{Dobbs08}, regulating star formation efficiency \citep{Tabatabaei18, Krumholz19}, and driving large-scale outflows that connect the disk with the CGM (e.g., \citealt{Mora-Partiarroyo19, Heald22, Li22, Li24a, Li24b, Lu23, Yang24}). Therefore, they contribute to the overall galactic ecosystem by mediating energy exchange between different phases of the ISM and providing a potential mechanism to sustain galactic winds \citep{Beck15, Han17}. The study of magnetic fields and CRs is essential for a comprehensive understanding of how galaxies evolve, how matter cycles between disks and halos, and how feedback processes shape the surrounding environment.

Edge-on galaxies provide ideal laboratories for investigating the connection between galactic disks and their surrounding halos. Their orientation allows the vertical structure of the ISM to be directly resolved, enabling us to measure how far magnetic fields and CRs extend above the mid-plane and how they interact with extraplanar gas. Observations over the past decades have revealed that many edge-on spiral galaxies exhibit radio continuum emitting halos extending several kiloparsecs (kpc) above the galactic disk \citep[e.g.,][]{Wiegert15, Krause18, Irwin19b, Irwin24, Heesen25}. These radio halos are dominated by synchrotron emission in the GHz band by CR electrons spiraling around magnetic field lines \citep{Vargas18,Irwin24b}, which makes radio observations a powerful probe of both the magnetic field configuration and the mechanisms of cosmic ray transport. The combination of polarization measurements and rotation measure (RM) synthesis allows us to reconstruct the 3D structure of magnetic fields \citep[e.g.,][]{Mora-Partiarroyo19b, Stein20, Xu25}. 

Advancing our understanding of galactic magnetic fields, therefore, requires systematic observations of a large sample of edge-on systems. The Continuum Halos in Nearby Galaxies—an EVLA Survey \citep[CHANG-ES;][]{Irwin12, Irwin12b} was designed to meet this need by targeting 35 nearby edge-on galaxies with broadband radio continuum observation in $L$-, $S$-, and $C$-band. The full polarization data provided by this survey enable the detailed examination of magnetic field tomography through RM synthesis. By employing a stacking of the $C$-band polarization maps of 28 CHANG-ES galaxies, \cite{Krause20} detected a clear X-shaped structure of the magnetic field in the stacked image. \cite{Stein25} also found that a large fraction (11/18) of galaxies with extended polarized halo exhibit a common X-shape pattern. 

Among the CHANG-ES sample, NGC~4565 stands out as an intriguing counterexample. Located at a distance of 11.9 Mpc \citep{Wiegert15}, this massive edge-on spiral galaxy is characterized by a relatively low star formation rate of ${\rm SFR}\sim 0.96 \,M_{\odot}\,{\rm yr}^{-1}$ and, in particular, the second lowest SFR surface density of $0.94\times10^{-3}\,M_{\odot}\,{\rm yr}^{-1}\,{\rm kpc}^{-2}$ compared to other CHANG-ES sample galaxies \citep{Vargas19}. Previous radio continuum work in the $C$- and $L$-band have shown that this galaxy lacks a prominent halo \citep{Schmidt19}; correspondingly, its magnetic field is dominated by a disk-parallel component, in contrast to the X-shaped fields seen in more active systems \citep{Stein25}. This distinction renders NGC~4565 an important outlier in the CHANG-ES sample, providing a potential lower limit on the conditions required for the development of extended radio halos and vertical magnetic field components. 
In this work, we present new full-polarization $S$-band observations of NGC~4565. The $S$-band (2--4 GHz) is particularly powerful for Faraday tomography studies. It occupies a key frequency range that bridges the Faraday-thin regime, where higher-frequency C-band data have limited resolution in Faraday space, and the Faraday-thick regime, where lower-frequency L-band data often suffer from stronger Faraday depolarization. Therefore, our new data enable a more detailed and robust analysis of the magnetic field structure and cosmic ray transport than ever in NGC~4565.

This paper is organized as follows. Section \ref{sec:data} describes the observations and data reduction, including imaging and RM synthesis procedures. In Section \ref{sec:result}, we present the main results, covering the morphology of the radio continuum emission, magnetic field structures, vertical scale heights, and cosmic ray transport analysis. Section \ref{sec:discussion} discusses these findings in a broader context of galactic halos and magnetic fields, and Section \ref{sec:summary} summarizes our conclusions.

{\section{Observations and Methodology}\label{sec:data}

\subsection{VLA $S$-Band}

The $S$-band data, covering the frequency range $2\text{--}4~{\rm GHz}$, were obtained using 27 antennas of the Karl G. Jansky Very Large Array (VLA) in the C-configuration with full polarization. Observations were taken on 2021 July 23 through program 21A-033, as a follow-up survey of the original CHANG-ES survey in $C$- and $L$-bands \citep[PI: Y. Stein;][]{Irwin12}. The initial 2~GHz bandwidth of the $S$-band is divided into 16 spectral windows, each containing 64 channels. The largest angular scale in the VLA $S$-band C-configuration is $8\farcm2$, whereas the apparent angular size of NGC 4565 is larger \citep[$d_{25} = 16\farcm2$;][]{Irwin12}. To accommodate the extended emission of the galaxy, the observations were designed as a two-pointing mosaic, with the pointings placed symmetrically along the major axis at offsets of $\pm3\farcm75$ from the galactic center. The on-target integration time is 188 minutes per pointing, totaling 376 minutes for the entire mosaic. Nevertheless, large-scale structures exceeding the LAS of individual pointings ($> 8\farcm2$) may inherently be filtered out, implying that some missing flux is possible. 

Data reduction was carried out using version 6.6.4 of the Common Astronomy Software Applications package \citep[CASA;][]{CASA22}. The source 3C286 serves as the primary calibrator to determine the flux density scale, bandpass, and polarization angle. J1221+2813 is utilized as the secondary calibrator, while J1407+2827 acts as the zero-polarization calibrator to solve for instrumental polarization. Standard calibration procedures were applied, followed by two rounds of phase-only self-calibration before the final imaging. For the Stokes $I$ image, we employed the Multi-Scale Multi-Frequency Synthesis (MS-MFS) algorithm \citep{mtmfs}, using two Taylor terms and Briggs weighting with a robust parameter of 0. To account for the joint imaging of the mosaic pointings, the \texttt{gridder} parameter was set to \texttt{awproject} \citep{Bhatnagar08}. A systematic calibration uncertainty of 5\% is adopted here.

The resulting Stokes $I$ map, at a reference frequency of $3.0\,{\rm GHz}$, has a synthesized beam of $4\farcs99 \times 4\farcs62$, with an rms noise level of $3.5\,\mu{\rm Jy\,beam}^{-1}$. We also produced a tapered Stokes $I$ image by applying a Gaussian taper to $10\arcsec$, yielding a beam size of $10\farcs32 \times 10\farcs26$ and an rms noise of $3.8\,\mu{\rm Jy\,beam}^{-1}$. All these maps have been corrected for the primary beam response.

\subsection{Rotation Measure Synthesis} \label{sec:RMsyn}

The orientation of linearly polarized light, described by its polarization angle (PA or $\psi$), is altered as it passes through a magnetized plasma. This phenomenon, known as Faraday rotation, results in a change in the observed PA given by $\Delta \psi = \psi_{\rm obs} - \psi_{\rm int} = {\rm RM}\,\lambda^2$, where $\lambda$ is the observing wavelength and RM is the Rotation Measure. This rotation is governed by the Faraday depth ($\phi$), defined as the integral of the electron density ($n_{\rm e}$) and the line-of-sight magnetic field component ($B_{\parallel}$) along the path $L$: $\phi=0.81 \int_{L}n_{\rm e}B_{\parallel}{\rm d}l$. Analyzing this effect, therefore, provides a powerful diagnostic for probing magnetic field structures and recovering the intrinsic PA of the source. The technique to extract this information from broadband polarization data is RM synthesis \citep{Burn96, RMsyn}.

In this work, we perform RM synthesis using the \texttt{RM-Tools} package\footnote{\url{https://github.com/CIRADA-Tools/RM-Tools}} \citep{RM-tools26, RM-Tools}. Following the procedure outlined in \citet{Xu25}, each spectral window in the $S$-band was divided into four segments (each with a bandwidth of roughly 26~MHz due to the flagging of the edge channels). Given our frequency coverage of 2--4~GHz, the theoretical FWHM resolution in Faraday depth space is $\delta \phi \approx 2.0 \times 10^2 {\rm\,rad\,m}^{-2}$. The largest continuous Faraday depth structure that can be resolved is $\mathrm{max}_{\mathrm{scale}} \approx 5.6 \times 10^2 {\rm\,rad\,m}^{-2}$, and the maximum observable RM is $|| \phi_{\mathrm{max}}|| \approx 1.0 \times 10^4 {\rm\,rad\,m}^{-2}$.

For each segment, the Stokes $Q$ and $U$ images were produced using \texttt{Briggs} weighting with a \texttt{robust} parameter of 2, which optimizes the sensitivity to recover extended faint emission. Due to the mosaic observation of NGC~4565, we generated images for each individual pointing and then combined them using the CASA task \texttt{linearmosaic}. All resulting images were convolved to a uniform beam size of $15\arcsec$ and corrected for the primary beam response. RM synthesis was then conducted on the combined cubes over a Faraday depth range from $-2048 {\rm\,rad\,m}^{-2}$ to $+2048 {\rm\,rad\,m}^{-2}$ with a sampling spacing of $4 {\rm\,rad\,m}^{-2}$.

The polarized intensity (PI) and RM maps were constructed by fitting the amplitude and the exact location of the peak in the Faraday dispersion function (FDF) at each spatial pixel. A Ricean bias correction was applied to the peak PI values using ${\rm PI}_{\rm eff}=\sqrt{{\rm PI}_{\rm peak}^{2}-2.3\sigma_{\rm FDF}^2}$ \citep{Pol_bias}, where $\sigma_{\rm FDF}$ is the estimated noise in the PI Faraday spectrum, with a mean value of $\sim 3.12\,\mu{\rm Jy\,beam}^{-1}$. To obtain the intrinsic PA, the Stokes $Q$ and $U$ components at the FDF peak were interpolated from adjacent samples. The observed PA at the 2.95~GHz reference frequency was calculated from these components and then corrected using the RM values. The uncertainties for the resulting PI, PA, and RM maps are also derived during the peak-fitting process. Complete details regarding these uncertainty estimations can be found in Section 3.5 of \citet{RM-tools26}.

Importantly, this work focuses on isolating the single most dominant emission component along the line of sight, without modeling secondary Faraday structures. For this specific goal, the $S$-band alone provides an optimal balance between spatial resolution and Faraday depolarization, making it sufficient for our current analysis. While the CHANG-ES project offers full $L$-, $S$-, and $C$-band coverage, the dense disk is severely depolarized at the $L$-band, where its inclusion would primarily introduce noise rather than reliable polarization signals. Furthermore, robustly combining $S$- and $C$-band data for wide-band RM synthesis requires extensive methodological testing that is currently underway (e.g., N. Pourjafari et al., in prep.).

Finally, the RM map presented in this work has been corrected for the Galactic foreground contribution. To satisfy the divergence-free condition of the magnetic field, we adopted the error-weighted mean RM across the galaxy as the foreground estimate, yielding a subtraction of ${\rm RM}_{\rm fg} = -4.9 \pm 1.7{\rm\,rad\,m}^{-2}$.

\subsection{Other Multiwavelength Data}

To complement our $S$-band radio continuum observations so as to derive the spectral index, magnetic field strength, and CR transport parameters, we incorporate additional multiwavelength data sets to this work.

To perform the multi-band spectral index analysis, we use the 144~MHz image from the Data Release 3 (DR3) of the LOFAR Two-meter Sky Survey \citep[LoTSS;][]{LoTSS, LoTSS_DR3} with an angular resolution of $20\arcsec$ to better recover extended emission. We also incorporate the $L$-band data (combined B, C, and D configurations, 1.57~GHz) from \citet{Schmidt19} with an angular resolution of $12\arcsec$. To facilitate a direct comparison across all frequencies, we smoothed both our $S$-band image and the $L$-band image to match the $20\arcsec$ resolution of the LOFAR data. The combination of the 144~MHz, $L$-band, and $S$-band maps allows us to derive spatially-resolved spectral index distributions and conduct subsequent CR transport modeling.

To investigate the connection between the magnetic field structure and the multiphase ISM (see Section~\ref{sec:Mag_geometry}), we also employ \ion{H}{1} 21~cm line and soft X-ray data. The \ion{H}{1} datacube is taken from the HALOGAS survey \citep{HImap} observed with the Westerbork Synthesis Radio Telescope (WSRT). We chose the low-resolution version ($44\arcsec\times34\arcsec$ beam, velocity resolution $4.12\,{\rm km\,s^{-1}}$) to recover the faint extended neutral gas beyond the galactic disk. 

We generated the X-ray diffuse-emission image using the standard \emph{Chandra}/ACIS reduction pipeline. Two observations (ObsIDs 404 and 3950; total cleaned exposure $\simeq62$~ks) were reprocessed with CIAO and the latest CALDB; soft-proton flares were filtered and astrometry was aligned across ObsIDs. Science images and exposure maps were produced in the $0.5$--$1.5$~keV band, reprojected to a common tangent point, and merged into an exposure-corrected count-rate image. The instrumental (particle) background was removed using the stowed background scaled to the $9.5$--$12$~keV count rate. No astrophysical sky-background subtraction was applied, as this product is intended for visualization.
Point sources were detected with \texttt{wavdetect} and masked with energy-dependent PSF radii; mask handling and local inpainting follow the procedure described in \citet{Luan2025}. The merged, background-subtracted, exposure-corrected image was then lightly adaptively smoothed (display only; no impact on quantitative analyses) and delivered for overlay with the radio data.

For spectral extraction, we followed the workflow in \citet{Luan2025}. Diffuse-emission spectra were extracted from source-masked regions, with responses generated using CIAO tools. The instrumental background was constructed with the public software of \citet{Suzuki2021}, and the astrophysical background was accounted for by drawing realizations from the best-fit model to a large, source-free region located well away from NGC~4565. All X-ray spectral fittings were performed using XSPEC \citep{XSPEC1996}. The model names and parameter names mentioned in this paper are also the same as those in XSPEC.

\section{Results}\label{sec:result}

\subsection{Morphology of the Total Radio Continuum Emission}\label{sec:TotalIntensity}

\begin{figure*}[th!]
    \includegraphics[width=\linewidth]{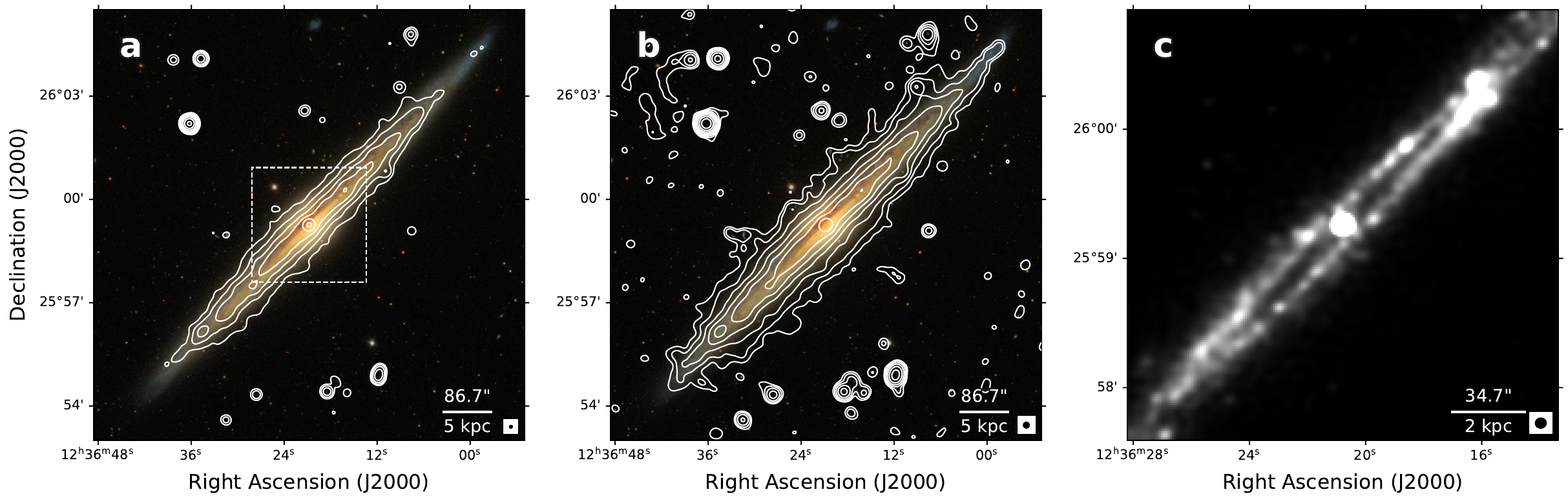}
    \caption{$S$-band total intensity images of NGC 4565. (\textit{a}) Contours from the robust zero weighting image are overlaid on a three-color SDSS optical image constructed from the $i$, $r$, and $g$ filters. The contour levels start at $3\sigma = 10.5 \,\mu{\rm Jy\,beam}^{-1}$ and increase by factors of 2. The beam size of this panel is $4\farcs99 \times 4\farcs62$. The white box indicates the region shown in panel (c). (\textit{b}) Contours from the $10\arcsec$ uv-tapered image, overlaid on the same SDSS optical image. The contours begin at $3\sigma = 11.2 \,\mu{\rm Jy\,beam}^{-1}$ and increase by factors of 2. The beam size of this panel is $10\farcs32 \times 10\farcs26$. (\textit{c}) A detailed grayscale view of the central ring-shaped structure from the robust zero-weighting image. In each panel, the synthesized beam is shown as a black ellipse in the bottom-right corner.}
    \label{fig:total_intensity}
\end{figure*}

Figure \ref{fig:total_intensity} shows the $S$-band C-array total intensity maps of NGC~4565. The radio continuum emission is largely confined to the galactic disk. In contrast to the other galaxies in the CHANG-ES sample, NGC~4565 lacks a prominent, vertically extended radio halo. Even in the uv-tapered map, which is more sensitive to large-scale features, no sign of such an extended halo is found. The radial and vertical extents of the emission, measured from the $3\sigma$ contour of the uv-tapered map, are $\sim 24.4$~kpc and $\sim 4.2$~kpc, respectively. The resulting radial-to-vertical extent ratio of $\sim 6$ is significantly larger than the ratio of 1--2 found by \cite{Wiegert15} in their  $L$-band stacking analysis of 30 CHANG-ES galaxies. The total flux density enclosed by the $3\sigma$ contour of the uv-tapered map is $79.4 \pm 4.0 {\rm\,mJy}$. After subtracting the contribution from the central core ($\sim 5.0{\,\rm mJy}$, see details below), the flux density is $\sim 74.4{\,\rm mJy}$. This value falls squarely between the CHANG-ES $L$-band ($139 \pm 7 {\rm\,mJy}$) and $C$-band ($39.5 \pm 5.9 {\rm\,mJy}$) measurements \citep{Schmidt19}, as expected for the intermediate frequency.

At the galactic center, NGC~4565 hosts a compact radio core, consistent with the presence of a potential AGN \citep{Irwin19a,Lambrides19}. By fitting this core to a 2-D Gaussian using the CASA task \texttt{imfit} within a region twice the size of the synthesized beam, we find a flux density for the core of $5.0 \pm 0.3 {\rm\,mJy}$.
The central region of NGC~4565 is also featured by a prominent ring, previously observed in dust emission \citep{Laine10, Kormendy10}, molecular gas \citep{Neininger96, Yim14}, and $C$-band radio continuum \citep{Schmidt19}. This ring, which is similar to the radio nuclear ring in NGC~5792 \citep{Yang22}, is clearly resolved in our $S$-band C-array data (Figure \ref{fig:total_intensity}c). We modeled this feature using the \texttt{GaussianRing3D} function in the \texttt{IMFIT} package \citep{IMFIT} to deproject its geometry, yielding an inclination of $85\fdg8^{+0.4}_{-0.8}$ and a ring radius of $5.6^{+0.3}_{-0.2} {\rm\,kpc}$.

\subsection{Magnetic Fields}\label{sec:mag_field}

Our linear polarization observations, combined with the RM synthesis technique, provide a comprehensive view of the magnetic field in NGC~4565 with its components in the plane of the sky and along the line of sight (LOS) both recovered. Figure \ref{fig:magnetic_field} demonstrates the resultant polarization properties.

\begin{figure*}[th!]
    \includegraphics[width=\linewidth]{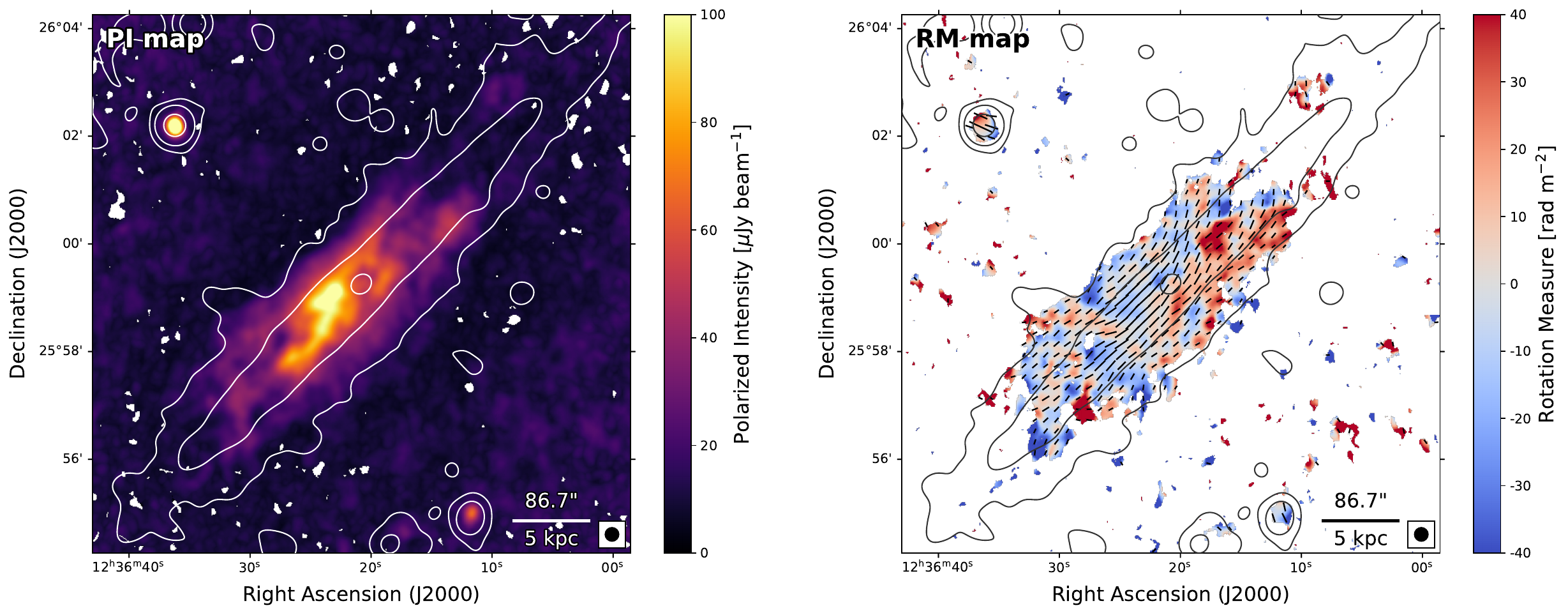}
    \caption{Polarization properties of NGC~4565 derived from RM synthesis. \textit{Left panel}: Map of the polarized intensity. \textit{Right panel}: RM map, with the Galactic foreground contribution subtracted. The intrinsic magnetic field orientations, corrected for Faraday rotation, are shown as black lines. The RM values and field lines are blanked for polarized intensities below $5\sigma$, where $\sigma = 3.12 \,\mu{\rm Jy\,beam}^{-1}$. In both panels, total intensity contours from Figure \ref{fig:total_intensity}b at levels of 3, 24, and 192$\sigma$ are overlaid to outline the galaxy's structure. The $15\arcsec$ synthesized beam is displayed in the bottom-right corner.}
    \label{fig:magnetic_field}
\end{figure*}

The polarized intensity (PI) in NGC~4565 is widespread, with a spatial extent comparable to that of the total intensity halo. The distribution exhibits a remarkable asymmetry with respect to the minor axis, with the bulk of the polarized emission found on the southeast side. The PI also peaks at a distance of $\sim3$ kpc southeast of the nucleus. This asymmetry, also seen in the $L$-band and $C$-band \citep{Sukumar91, Schmidt16}, is likely a consequence of lower Faraday depolarization along the sight lines penetrating the part of the galaxy as a foreground (in this case, the southeast side; \citealt{Zheng22}), where the path through the magneto-ionic medium is shorter.

The magnetic field structure is detailed in the right panel of Figure \ref{fig:magnetic_field}. The RM traces the LOS component of the magnetic field, while the orientations, derived by rotating the polarization angles by $90^{\circ}$, represent the component in the plane of the sky.
After correcting for the Galactic foreground contribution (${\rm RM}_{\rm fg} = -4.9 \pm 1.7{\rm\,rad\,m}^{-2}$; see Section \ref{sec:RMsyn}), the RM values in NGC~4565 are mostly (95\%) within the range of $\pm 50 {\rm\,rad\,m}^{-2}$, with a mean uncertainty of $\sim 10 {\rm\,rad\,m}^{-2}$. Positive and negative values correspond to a magnetic field pointing toward and away from the observer, respectively. 
Overall, the RM distribution exhibits a complex spatial variation. The RM profile along the major axis (Figure \ref{fig:rm_profile}) reveals an asymmetric pattern across the disk. Notably, there is a local RM maximum at $\sim 4.5 {\rm\,kpc}$ (NW) and a local minimum at $\sim -7 {\rm\,kpc}$ (SE). These locations spatially correlate with the edge of the radio ring structure (Figure \ref{fig:total_intensity}c), suggestive of a magnetic field structure closely coupled with the ring. 
This pattern of sign reversal across the nucleus is consistent with the projection of an axisymmetric spiral magnetic field viewed edge-on \citep{Beck15}.

\begin{figure}
    \centering
    \includegraphics[width=\linewidth]{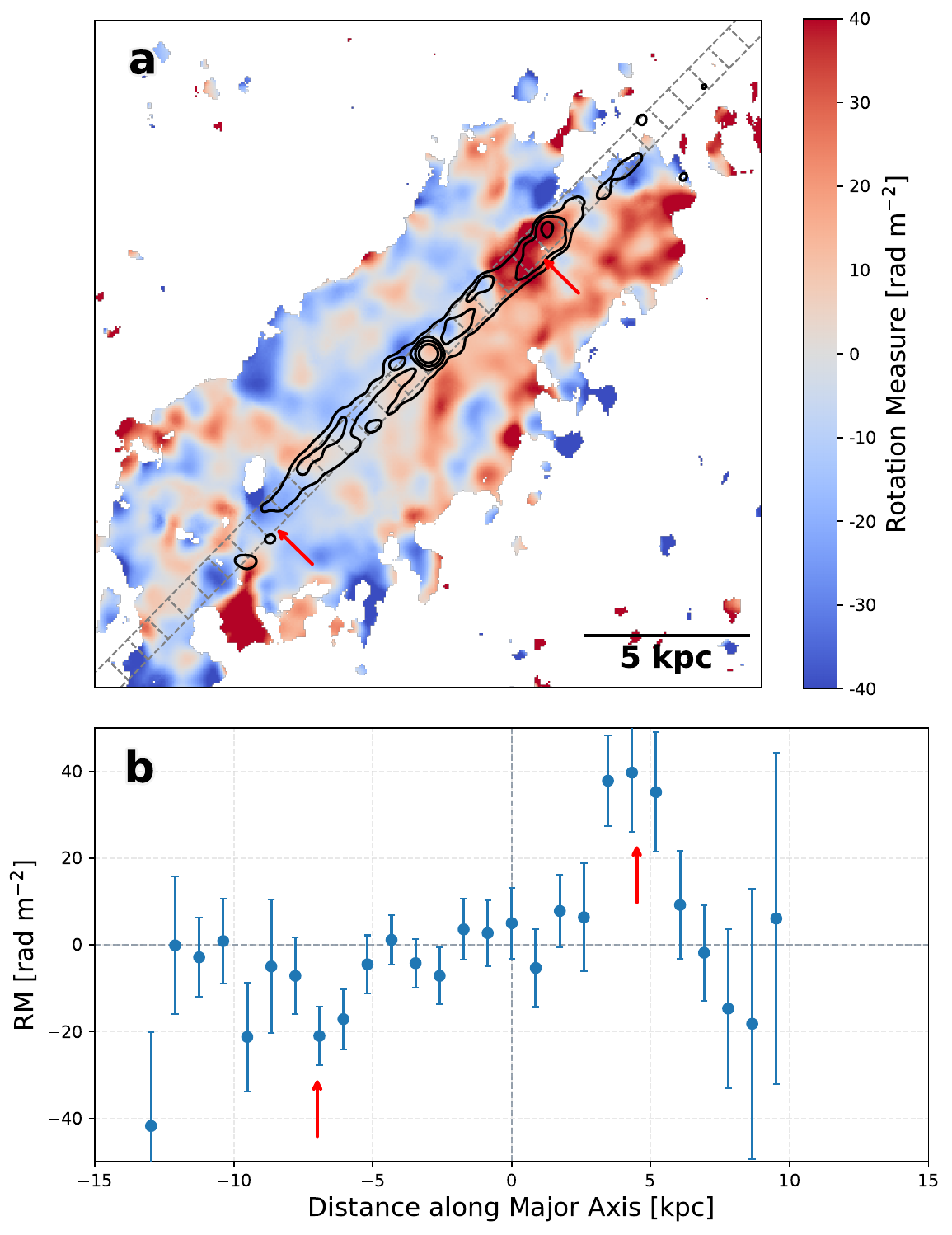}
    \caption{RM distribution along the major axis of NGC~4565. (\textit{a}) The RM map overlaid with $S$-band total intensity contours (black lines, levels at 30, 40, 60, and $120\,\sigma$; see Figure \ref{fig:total_intensity}c) to highlight the inner radio ring structure. The rectangular boxes indicate the $15\arcsec\times15\arcsec$ bins used to extract the RM profile. (\textit{b}) The RM profile along the major axis. Each data point represents the inverse-variance weighted mean RM within the bin. The $x$-axis represents the distance from the galactic center, where negative values correspond to the southeast side. In each panel, the red arrows mark the positions of the local maximum ($\sim 4.5$ kpc, NW) and local minimum ($\sim -7$ kpc, SE).}
    \label{fig:rm_profile}
\end{figure}

Regarding the magnetic field component in the plane of the sky, previous work at $L$- and $C$-bands revealed a disk-parallel pattern, with field orientations predominantly aligned with the major axis \citep{Sukumar91, Krause20, Stein25}. In particular, the systematic analysis by \cite{Stein25} of 18 CHANG-ES galaxies classified NGC~4565 as an archetypal disk-dominated system, in contrast to the X-shaped configurations seen in many other edge-on galaxies. While our $S$-band data confirm the strong disk-parallel pattern near the galactic plane, they also reveal magnetic field lines pointing away from the plane on the northeast side of the galaxy. 
Indication of a similar feature is seen in the lower-resolution $L$-band D-array data from \citet{Schmidt19}. The increased angular resolution of our new S-band observations has made it feasible to trace this structure in greater detail, revealing an organized field turning away from the disk.

To quantify this trend, we partitioned the galaxy into four quadrants as per its minor and major axes \citep[position angle $135\fdg5$;][]{Schmidt19}, following the method of \cite{Stein25}. The magnetic field orientations ($\chi$), which ranges from $0^{\circ}$ to $180^{\circ}$, is a circular quantity with identical endpoints. Obviously, standard linear statistics are inappropriate for such data. For instance, the arithmetic mean of $170^{\circ}$ and $10^{\circ}$ is $90^{\circ}$, which incorrectly implies a perpendicular orientation, whereas the true circular mean is $0^{\circ}$ (or $180^{\circ}$ ), indicating a parallel orientation. 
Therefore, to properly analyze their distribution in each quadrant, we employ circular statistics \citep{circstat} using the \texttt{PyCircStat2} package\footnote{\url{https://github.com/circstat/pycircstat2}}. Results are presented in Figure \ref{fig:PA0_distribution} and Table \ref{tab:PA0_distribution}. Following the classification of \cite{Stein25}, a disk-dominated pattern is characterized by $\chi$ close to $0^{\circ}/180^{\circ}$ in all quadrants, whereas an X-shaped pattern is expected to have $10^{\circ}<\chi<90^{\circ}$ in quadrants I \& III and $90^{\circ}<\chi<170^{\circ}$ in quadrants II \& IV.

Our analysis shows that quadrants III and IV (the southwest side) exhibit a prominent disk-dominated pattern, with circular means deviating from the major axis by less than $10^{\circ}$. In contrast, the circular mean in quadrant I is $26\fdg5\pm0\fdg4$, which clearly agrees with the definition of an X-shaped pattern. The case in Quadrant II is of higher complication, with a circular mean of $171\fdg6\pm0\fdg2$. Its rose diagram reveals a bimodal distribution, with one peak aligned with the disk (near $180^{\circ}$) and the other peak consistent with an X-shaped field (less than $170^{\circ}$). We therefore conclude that the northeast side of the galaxy (quadrants I and II) exhibits strong evidence for an X-shaped magnetic field. The trend is particularly evident in quadrant I. 

\begin{figure}[ht!]
    \includegraphics[width=\linewidth]{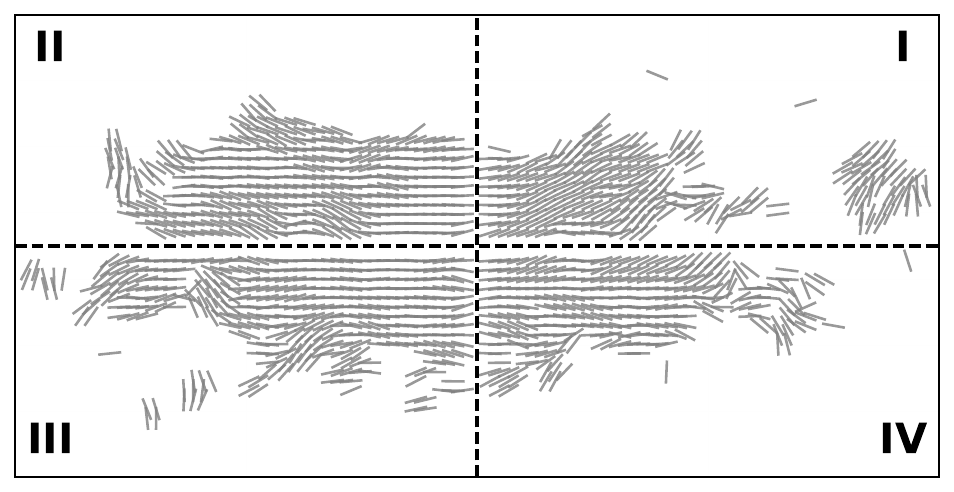}
    \includegraphics[width=\linewidth]{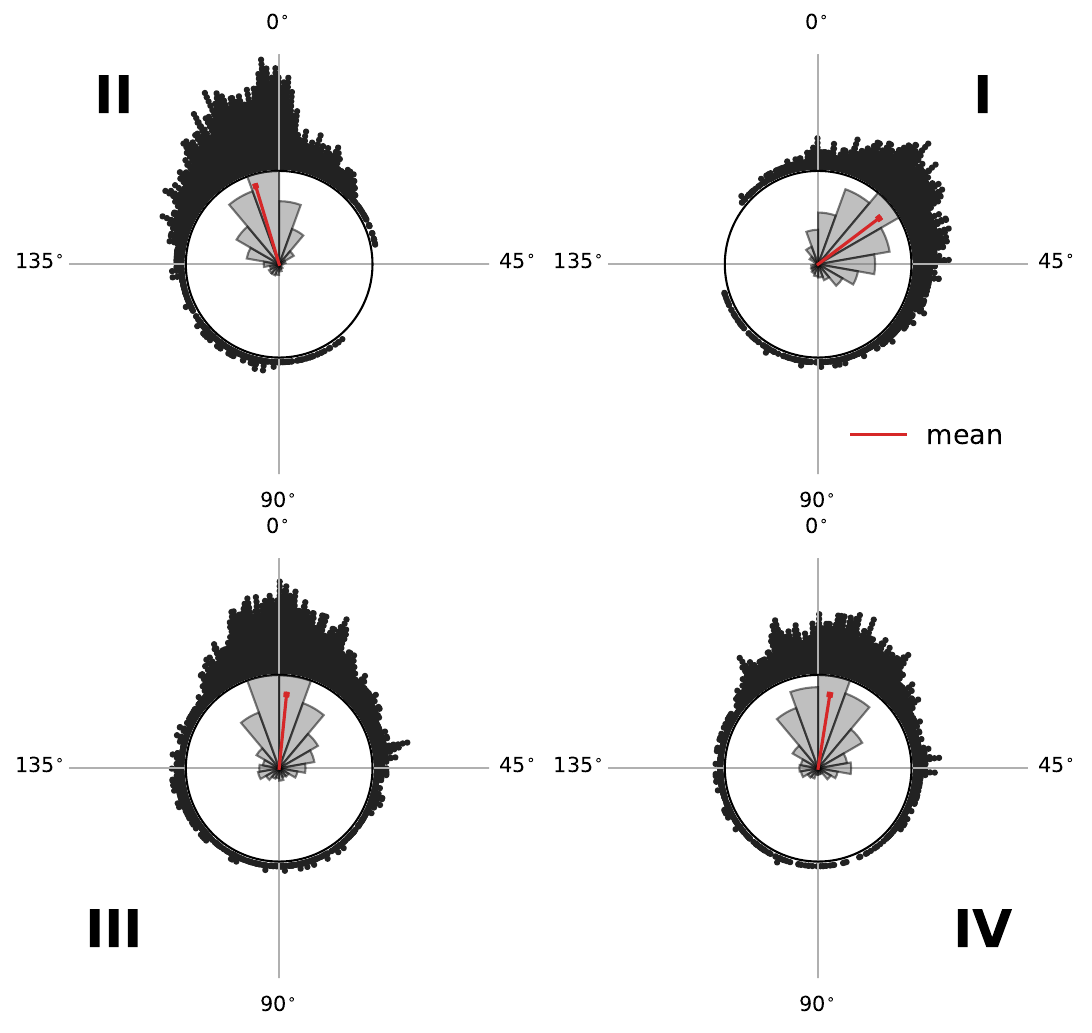}
    \caption{Distribution of magnetic field orientations in four galactic quadrants. (\textit{Top}) Map of the magnetic field orientations, where the data from Figure \ref{fig:magnetic_field} has been divided into four quadrants (labeled I, II, III, and IV). Regions around the galactic major and minor axes have been masked out. (\textit{Bottom}) Rose diagrams illustrating the magnetic field orientation distribution for each quadrant. In each plot, individual data points are scattered around the circle, and the angular distribution is shown as a gray histogram. The solid red line indicates the circular mean of the distribution.}
    \label{fig:PA0_distribution}
\end{figure}

\begin{deluxetable}{lcc}[ht]
    \tablecaption{Circular Statistics of Magnetic Field Orientations by Quadrant}
    \label{tab:PA0_distribution}
    \tablehead{
        \colhead{Quadrant} & \colhead{Circular Mean} & \colhead{Circular Standard Deviation}\\
        \colhead{} & \colhead{$(^{\circ})$} & \colhead{$(^{\circ})$}
    }
    \startdata
    I & $26.5\pm0.4$ & $17.3$ \\
    II & $171.6\pm0.2$ & $14.3$ \\
    III & $2.9\pm0.3$ & $18.4$ \\
    IV & $4.5\pm0.4$ & $18.2$ \\
    \enddata
    \tablecomments{The uncertainties quoted for the circular mean represent the 95\% confidence interval. The circular standard deviation is a measure of the angular spread around the mean.}
\end{deluxetable}

In summary, our sensitive $S$-band observations confirm the dominant disk-parallel magnetic field in NGC~4565 but also reveal, for the first time, quantitative evidence of an underlying X-shaped magnetic field component in the halo on the northeastern side of the galaxy. A more detailed discussion of these findings is provided in Section \ref{sec:discussion}.

\subsection{Vertical Scale Heights}\label{sec:ScaleHeight}

\begin{figure*}[ht]
    \includegraphics[width=\linewidth]{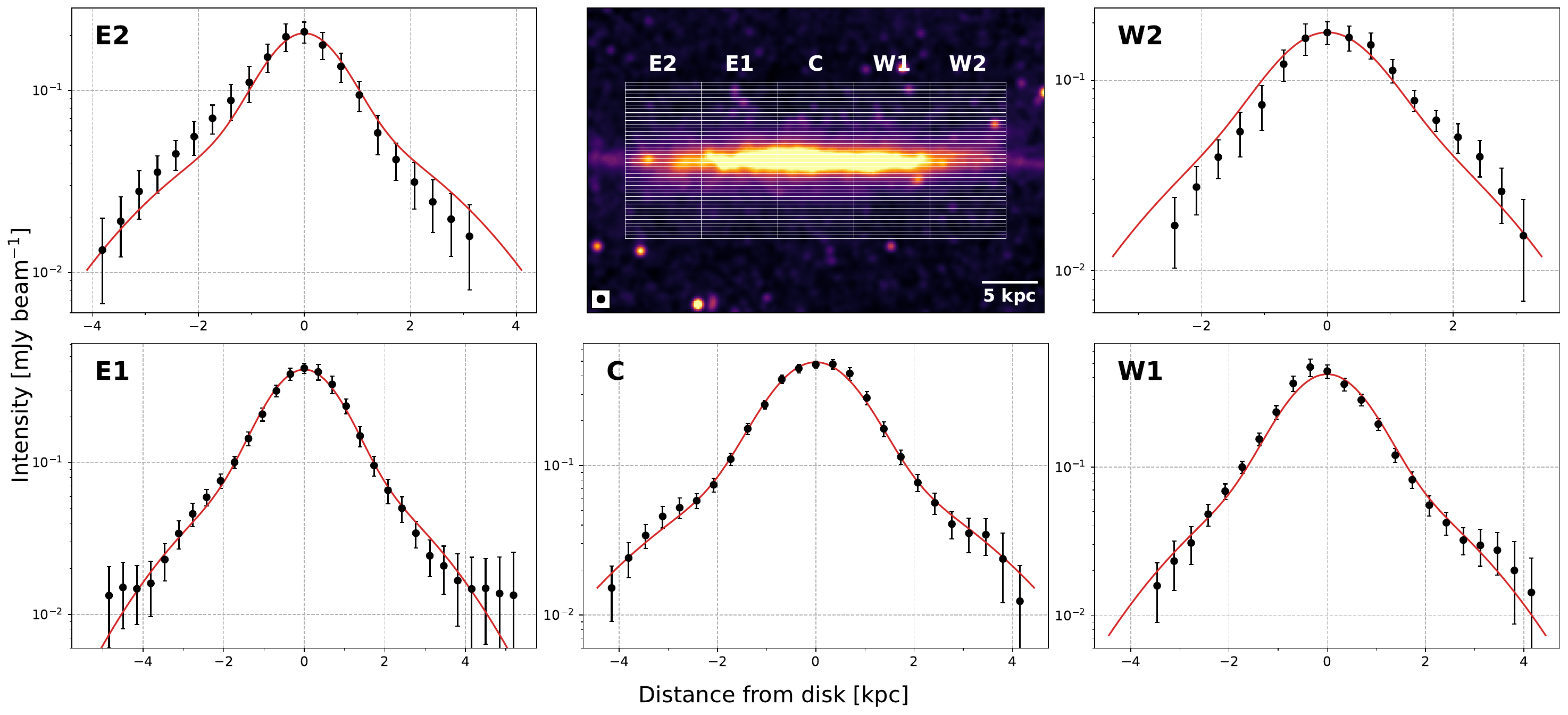}
    \caption{Vertical profiles of the non-thermal radio continuum emission at $S$-band. The top-center panel shows the locations of the five analysis strips (E2, E1, C, W1, and W2) overlaid on the non-thermal intensity map. The other five panels display the corresponding vertical profile for each strip, arranged symmetrically around the setup panel. In each profile panel, the black points represent the measured flux density, averaged within $6\arcsec\times120\arcsec$ ($0.35\times6.92$ kpc) boxes. Positive distances are to the north of the major axis, and negative distances are to the south. The red curve is the best-fit two-component Gaussian profile.}
    \label{fig:profile}
\end{figure*}

To quantify the vertical extent of the non-thermal radio emission in the galactic halo, we fit the vertical intensity profiles, following the procedures adopted in previous works \citep{Dumke95, Krause18, Stein23}. The fitting was conducted on a non-thermal intensity map, which we created by subtracting the thermal emission from a smoothed, $12\arcsec$ resolution uv-tapered image. The thermal emission map was adapted from \cite{Schmidt19}, who estimated the thermal contribution from a combination of H$\alpha$ and \textit{Spitzer} $24\,\mu{\rm m}$ emission. In $S$-band, the thermal fraction ranges from $27\% \pm 13 \%$ in the disk to $4\% \pm 2\%$ in the halo. Here, a conservative relative uncertainty of 50\% is adopted for the thermal estimates based on the value reported in \citet{Schmidt19}. We used the \texttt{BoxModels} task within the \texttt{NOD3} package \citep{NOD3} to fit vertical intensity profiles in strips oriented perpendicular to the major axis.

We modeled the vertical profiles in each strip using two different two-component functions: a Gaussian, defined as $w_{\rm gauss}(z) = w_{0}\exp(-z^{2}/h^{2})$, and an exponential, given by $w_{\rm exp}(z) = w_{0}\exp(-|z|/h)$. To account for the beam size and the projection effects in inclined galaxies, these intrinsic profiles were convolved with a Gaussian kernel:
\begin{equation}
    g(z) = \frac{1}{\sqrt{2\pi\sigma^{2}}}\exp(-z^{2}/2\sigma^2),
\end{equation}
where $\sigma$ corresponds to the effective beam width, $\sigma = {\rm HPBW}_{\rm eff}/(2\sqrt{2\ln2})$. We adopted an inclination angle of $i = 85\fdg8$ for NGC~4565 (see Section \ref{sec:TotalIntensity}). Further details on this fitting procedure can be found in \citet{NOD3}. In our analysis, the exponential fits were unsuccessful, yielding relative uncertainties in the fitted parameters that exceeded 100\%. Consequently, we conclude that the vertical distribution of NGC~4565 is best described by the two-component Gaussian model, which we adopt for all subsequent analyses. 
The strip locations and the resultant best-fit profiles are shown in Figure \ref{fig:profile}, and the derived Gaussian scale heights for the narrower (hereafter, ``disk''; $h_{\rm disk}$) and broader (hereafter, ``halo''; $h_{\rm halo}$) Gaussian components are listed in Table \ref{tab:scale_height}.

Averaging the results from all strips gives mean Gaussian scale heights of $h_{\rm disk}=0.82\pm0.04 {\rm\,kpc}$ and $h_{\rm halo}=3.02\pm0.13 {\rm\,kpc}$. We note that \citet{Heesen25} derived significantly smaller exponential scale heights of $0.60\pm0.20\,{\rm kpc}$ (disk) and $0.83\pm0.06\,{\rm kpc}$ (halo) using the same dataset. This discrepancy is likely attributed to our use of uv-tapering, which enhances the sensitivity to large-scale diffuse structures compared to their high-resolution analysis. Moreover, since a Gaussian profile decays faster in the wings than an exponential one, a larger scale height is required to fit the same extended emission. Finally, our profiles are corrected for the thermal contribution, which may reduce the impact of the thin disk component.

The scale height is maximal in the central strip and decreases towards larger distances from the galactic center. We notice a significant asymmetry in the outermost strips (E2 and W2). In strip E2, the observed emission shows a systematic deviation from the best-fit profile, with an excess in the northern part and a deficit in the southern part. The opposite trend is observed in strip W2. This asymmetry is reminiscent of the known warp in the disk of NGC~4565, which has been detected in low-frequency radio continuum \citep[144 MHz;][]{Heesen19}, \ion{H}{1} \citep{Yim14}, and optical observations \citep{Gilhuly20}. While the warped structure is not noticeable in our $S$-band total intensity map, the observed asymmetry in the vertical profiles of the outer strips is suggestive of its influence on the radio halo.

\begin{deluxetable}{lcccc}[ht!]
    \tablecaption{Vertical Gaussian Scale Height Fitting Results for NGC~4565}
    \label{tab:scale_height}
    \tablehead{
    \colhead{Strip} & \colhead{r} & \colhead{$h_{\rm disk}$} & \colhead{$h_{\rm halo}$} & \colhead{$\chi^{2}_{\rm red}$} \\ 
    \colhead{}& \colhead{(kpc)} & \colhead{(kpc)} & \colhead{(kpc)} & \colhead{}
    }
    \startdata
    E2 & $-13.85$  & $0.83\pm0.17$  & $3.01\pm0.48$ & 0.86 \\
    E1 & $-6.92$  & $0.89\pm0.06$  & $2.98\pm0.18$ & 0.35 \\
    C & $0.00$  & $0.78\pm0.05$  & $3.22\pm0.22$ & 0.33 \\
    W1 & $6.92$  & $0.79\pm0.11$  & $2.69\pm0.37$ & 0.82 \\
    W2 & $13.85$  & $0.94\pm0.38$  & $2.54\pm1.02$ & 1.32 \\
    \hline
    \multicolumn{2}{c}{Average} & $0.82\pm0.04$  & $3.02\pm0.13$ & - \\
    \enddata
    \tablecomments{Parameters are derived from the two-component Gaussian fits to the vertical profiles in five strips across NGC~4565.
    $h_{\rm disk}$ and $h_{\rm halo}$ represent the Gaussian scale heights of the disk and halo components, respectively.
    $r$ is the distance from the galactic center along the major axis.
    The final row lists the inverse-variance weighted average of the parameters of the five strips.
    }
\end{deluxetable}



\subsection{Non-thermal Spectral Index}
\label{sec:spectral_index}

To derive the non-thermal spectral index ($\alpha_{\rm nth}$) distribution, we employ our non-thermal $S$-band 3.0~GHz map (from Section \ref{sec:ScaleHeight}), the $L$-band 1.57~GHz map from \citet{Schmidt19}, and the LOFAR 144~MHz map from \citet{LoTSS_DR3}. All these images were downgraded to a common angular resolution of $20\arcsec$ before comparison. In the same manner as for the $S$-band data, we subtracted the estimated thermal contribution from both the LOFAR and $L$-band maps. The thermal correction for the low-frequency LOFAR data is insignificant; the average thermal fraction is merely $4.8\% \pm 2.4\%$ in the disk and, negligibly, $0.2\% \pm 0.1\%$ in the halo. For the $L$-band, the thermal fraction is $18\% \pm 9\%$ in the disk and $1.3\% \pm 0.7\%$ in the halo. By fitting these three thermal-subtracted maps to a single power-law model ($I_{\nu} \propto \nu^{\alpha}$), we derived the spectral index map shown in Figure~\ref{fig:specindex}. We obtain a spectral index of $\alpha = -0.77 \pm 0.06$ in the galactic mid-plane, consistent with the typical galactic non-thermal spectral index of $\sim-0.8$ \citep{Condon92}. In the halo, the spectrum shows a moderate steepening to $\alpha = -0.98 \pm 0.08$. The overall vertical steepening trend is consistent with those found by \citet{Heesen19} (144~MHz--1.5~GHz) and \citet{Schmidt19} (1.5--6~GHz). This gradient supports a scenario in which young cosmic-ray electrons (CRe) are injected in the star-forming mid-plane and they age subsequently when being transported into the halo. We note that low-frequency free-free absorption might cause the spectral index to be partially overestimated (i.e., flatter), especially within the galactic mid-plane, without significantly affecting the steepening trend in the halo.

\subsection{Equipartition Magnetic Field Strength}
\label{sec:magnetic_field}

We estimate the magnetic field strength by assuming energy equipartition between the total CRs and the magnetic field, using the revised formula from \cite{Beck05}:
\begin{equation} \label{equ:magnetic_field}
    B_{\mathrm{eq}}=\left[\frac{4 \pi(1-2\alpha )(K_{0}+1)I_{\mathrm{nt}}E_{\mathrm{p}}^{1+2\alpha}(\nu/2c_{1})^{-\alpha}}{(-2\alpha -1 )c_{2}(-\alpha)l c_{4}}\right]^{\frac{1}{3-\alpha}}.
\end{equation}
Here, $\alpha$ is the non-thermal spectral index as derived above, $I_{\mathrm{nt}}$ is the $S$-band non-thermal intensity at a frequency of $\nu=3{\rm\,GHz}$, $E_{\mathrm{p}}$ is the proton rest energy, and the terms $c_1$ through $c_4$ are constants as defined in \citet{Beck05}. The parameter $K_0$ is the ratio of the number densities of CR protons to electrons, for which we assume a constant value of 100. The parameter $l$ is the path length through the synchrotron-emitting medium along the line of sight. We estimate it following \cite{Schmidt19}, who modeled the galaxy as an oblate spheroid with a semi-major axis $R$ and semi-minor axis $Z$, yielding a path length of $l = 2R\sqrt{1-(r/R)^2-(z/Z)^2}$. Rigorously, this formula is only valid for a galaxy with an inclination of exactly $90^{\circ}$, though the inclination of NGC~4565 is sufficiently close to edge-on ($\sim 86^{\circ}$) and the resultant uncertainty is insignificant. We found $R=25{\rm\,kpc}$ and $Z=5{\rm\,kpc}$ from the extent of our total intensity map. We note that this equipartition assumption is valid globally in normal star-forming galaxies and locally on scales larger than $\sim 1$~kpc \citep{Seta19}. Accordingly, we masked out the $20\arcsec$ region centered on the Seyfert nucleus to exclude its potentially anomalous contribution.

The final magnetic field strength map is presented in Figure \ref{fig:Bmap}. The average field strength is found to be $4.6 \pm 1.0\,\mu{\rm G}$ in the disk and $3.5 \pm 0.7\,\mu{\rm G}$ in the halo, where the uncertainties are estimated by conservatively assuming a 50\% uncertainty in both $K_{0}$ and $l$. 
In comparison, \cite{Krause18} reported average magnetic field strengths of $9\text{--}11\,\mu{\rm G}$ over their sample of 13 CHANG-ES galaxies, with disk strengths of $11\text{--}15\,\mu{\rm G}$. Hence, the magnetic field in NGC~4565 is remarkably weaker. 

\begin{figure}[ht!]
    \centering
    \includegraphics[width=\linewidth]{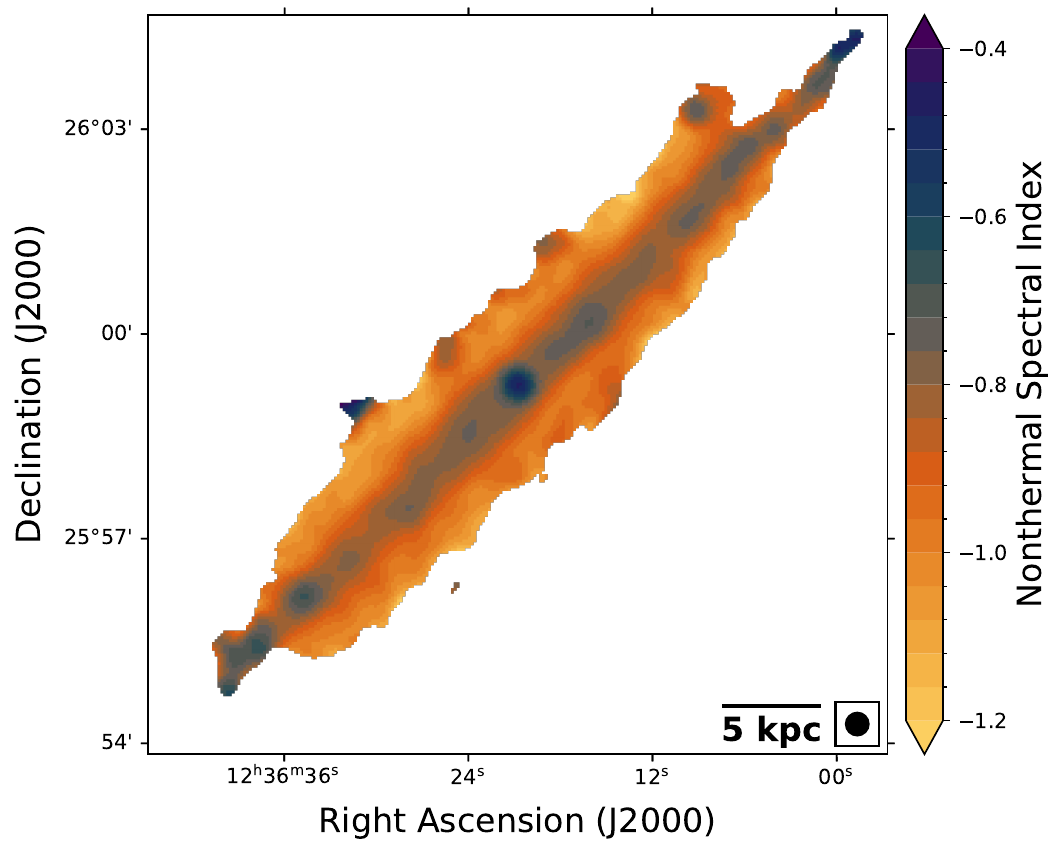}
    \caption{Distribution of the nonthermal spectral index derived by fitting a single power-law to the 144~MHz (LOFAR), 1.57~GHz (VLA $L$-band), and 3.0~GHz (VLA $S$-band) data. The map is visualized using the colormap from \cite{colormap}. Only regions with intensities $>3\sigma$ at all three frequencies are taken into account. The beam of $20\arcsec$ is displayed in the bottom-right corner. }
    \label{fig:specindex}
\end{figure}

\begin{figure}[ht!]
    \centering
    \includegraphics[width=\linewidth]{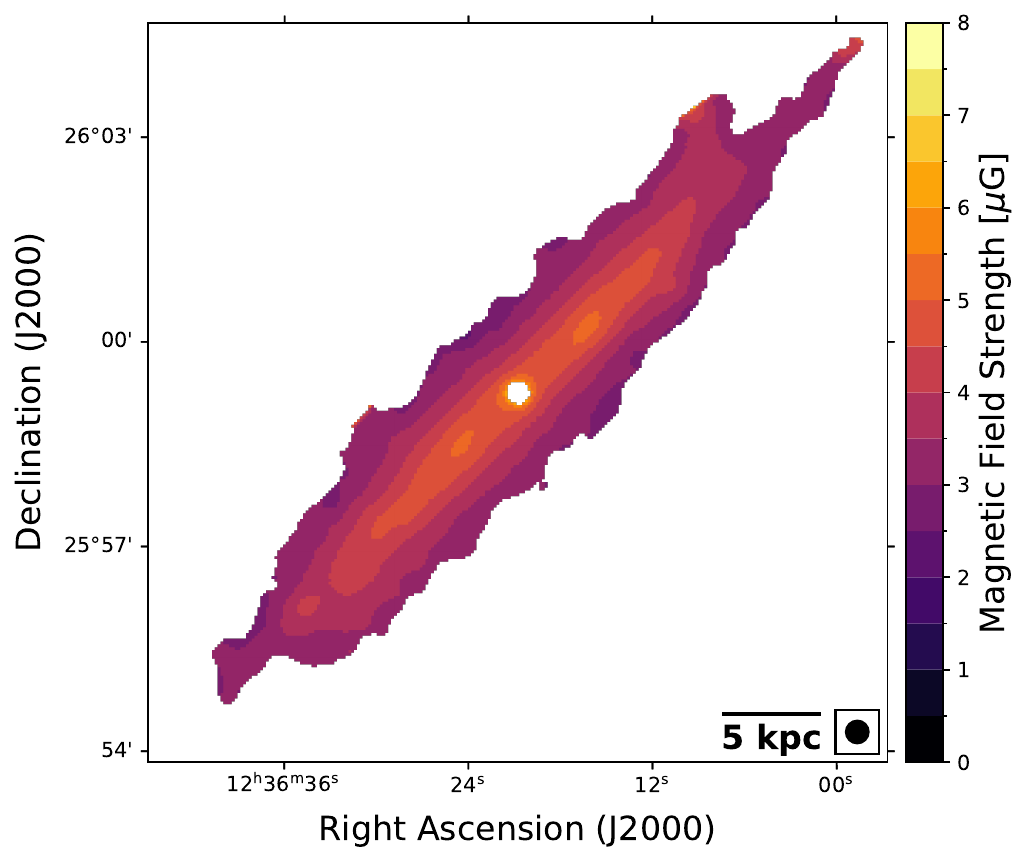}
    \caption{Magnetic field strength map of NGC~4565, computed assuming energy equipartition. The beam of $20\arcsec$ is displayed in the bottom-right corner.}
    \label{fig:Bmap}
\end{figure}

\subsection{Cosmic Ray Transport Mechanism}
\label{sec:CR_model}

We model the vertical CR transport using the 1D \texttt{SPINNAKER} code \citep{Heesen16, Heesen18}, which jointly fits the nonthermal intensity and spectral index profiles. The input data for the model were extracted from ten profiles, created by dividing each of the five strips (defined in a manner similar to Section \ref{sec:ScaleHeight}) into northern and southern segments. The extraction boxes were set to $7.5\arcsec\times120\arcsec$ to match the $20\arcsec$ resolution of the input maps.

We test two transport scenarios to model the vertical profiles. In both scenarios, CRe are injected in the mid-plane ($z=0$) with a power-law energy spectrum $N(E) \propto E^{-\gamma}$. The injection index $\gamma$ is related to the injection non-thermal spectral index by $\alpha_{\rm inj} = (1-\gamma)/2$. The first is an energy-independent diffusion model ($\mu = 0$), characterized by a diffusion coefficient $D_0$ and a single exponential magnetic field distribution, $B(z) = B_{0}\exp(-|z|/h_{B})$. 

The second is the flux-tube advection model described by \citet{Heald22}. In this model, CRs are advected vertically in a flow with a cross-sectional area defined as
\begin{equation}
    A(z) = A_{0}\left[1+\left(\frac{z}{z_{0}}\right)^{\beta}\right],
    \label{eq:flux_area}
\end{equation}
which corresponds to a hyperboloidal shape when $\beta = 2$. The flow radius is defined by $A = \pi r^{2}$, and the magnetic field strength is assumed to follow 
\begin{equation}
    B(r,v) = B_{0}\left(\frac{r}{r_{0}}\right)^{-1}\left(\frac{v}{v_{0}}\right)^{-1}.
    \label{eq:B_field}
\end{equation}
Here, $B_{0}$, $r_{0}$, and $v_{0}$ denote the mid-plane magnetic field strength, flow radius, and advection speed, respectively (see Section 4.3 and Appendix A in \citealt{Heald22} for details). 
For the advection model, we adopt an initial outflow radius $r_{0} = 6.92\,{\rm kpc}$ (corresponding to the strip width of $120\arcsec$) and a rotation velocity of $v_{\rm rot}=244{\rm\, km\,s}^{-1}$ \citep{HImap}. The free parameters for this model are $B_{0}$, $\gamma$, $z_{0}$, $\beta$, and the critical point wind speed $v_{c}$, while the critical height $z_{c}$ is a derived quantity. Given the limited vertical extent of the emission in NGC~4565, we fix $B_{0}$ and the injection index $\gamma$ to their average values measured in the central box of each strip for all models.

We determine the best-fit parameters by jointly minimizing the reduced chi-square ($\chi^{2}_{\rm red}$) from the simultaneous fits of three intensity profiles (LOFAR, $L$-band, and $S$-band) and two spectral index profiles ($\alpha_{S}^{\rm LOFAR}$ and $\alpha_{S}^{L}$). As an example, Figure \ref{fig:CR_Fit} displays the best-fit profiles for the northern part of strip E1. The best-fit parameters for all ten profiles are presented in Table \ref{tab:CR_Fit}. 
According to the total $\chi^{2}_{\rm red}$ values listed in Table \ref{tab:CR_Fit}, the advection model is consistently the preferred transport scenario across majority of the strips, with the exception of two strips (W1~S and W2~S) where the differences between the two models are marginal. While the pure diffusion model can adequately describe the 144~MHz emission, providing a comparable or even slightly better fit in some regions, it fails to reproduce the spectral steepening observed at spectral index profiles.

For the diffusion scenario, the fits yield an average diffusion coefficient of $D_{0} = 2.5 \times 10^{28}\,{\rm cm}^{2}{\rm \,s}^{-1}$, a value at the lower end of the typical range for spiral galaxies like the Milky Way ($D_0 \sim (3-5) \times 10^{28}\,{\rm cm}^{2}{\rm \,s}^{-1}$; \citealt{Strong07}). 
For the advection scenario, we find that the derived critical heights ($z_c$) for all strips exceed $5\,{\rm kpc}$ (see Table \ref{tab:CR_Fit}). This is beyond the vertical extent of the halo detected in the $S$-band, implying that the outflow within the observed vertical extent remains subsonic and has not yet reached the transition point to a supersonic galactic wind. The resulting velocity profiles for all strips are presented in Figure \ref{fig:v_profile}. We find a low average initial wind velocity at the mid-plane of $v_{0} \approx 60 {\rm\,km\,s}^{-1}$. The acceleration is gradual; at a height of $z=4\,{\rm kpc}$, the wind speed remains generally below $150 {\rm\,km\,s}^{-1}$. This advection speed is substantially lower than the galactic winds (typically several hundred ${\rm km\,s}^{-1}$) observed in other CHANG-ES galaxies \citep[e.g.,][]{Heesen18, Heald22, Stein23}. 

Furthermore, the outflow appears insufficient to escape the galactic potential. Based on the rotation velocity of $v_{\rm rot}=244{\rm\,km\,s}^{-1}$ \citep{HImap}, we estimate a conservative lower limit for the escape velocity of $v_{\rm esc}\approx \sqrt{2}v_{\rm rot}\approx 345{\rm\,km\,s}^{-1}$ (neglecting the dark matter halo). Since the wind velocity $v(z)$ remains significantly lower than $v_{\rm esc}$, the CRs and magnetized gas are likely to remain gravitationally bound. This evidence strongly supports the galactic fountain scenario previously suggested for NGC~4565 by \citet{Heesen19}, rather than a wind escaping into the intergalactic medium.

In summary, our fitting outcomes suggest that the advection model provides a superior description of the vertical CRe transport in NGC~4565 compared to the diffusion model. While diffusion may still contribute at lower frequencies, the advection model is required to explain the global spectral steepening. Furthermore, the derived low outflow velocities ($v \ll v_{\rm esc}$) indicate that the cosmic rays and magnetic fields likely form a confined galactic fountain rather than an escaping wind.

\begin{figure*}[ht!]
    \centering
    \includegraphics[width=0.48\linewidth]{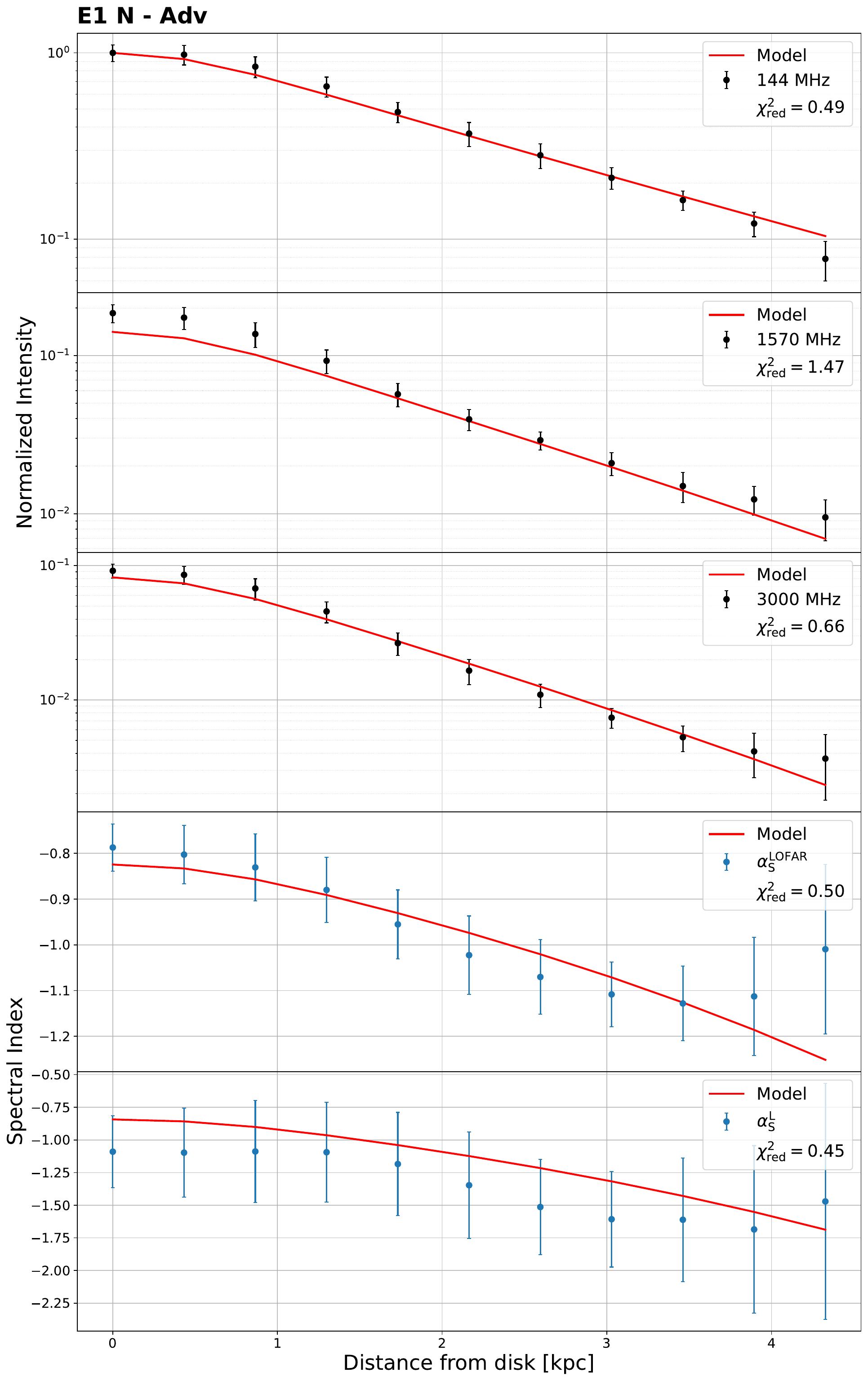}
    \includegraphics[width=0.48\linewidth]{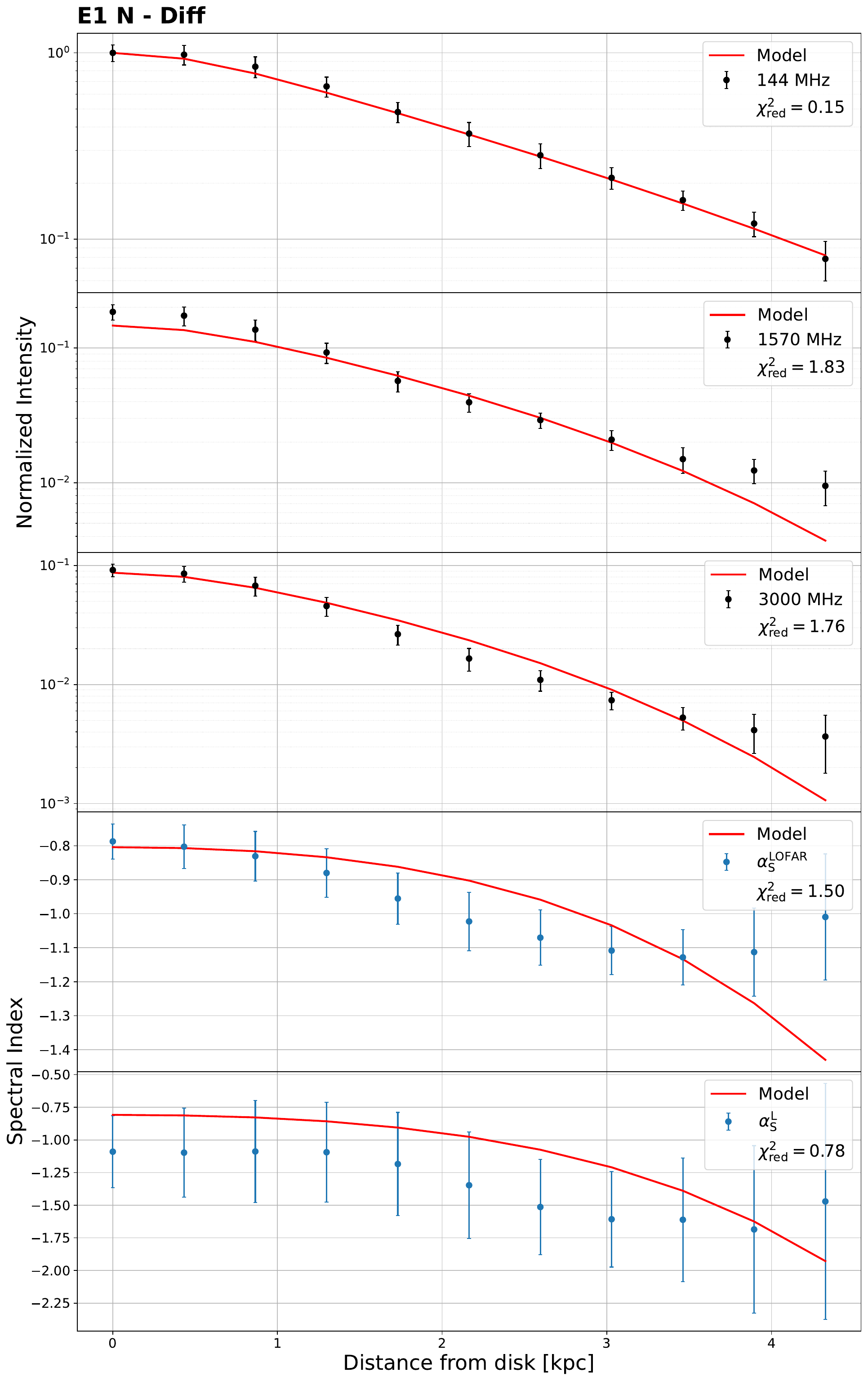}
    \caption{Best-fit advection (\textit{left panels}) and diffusion (\textit{right panels}) models in the northern part of strip E1. From top to bottom, the panels depict the \texttt{SPINNAKER} fits to the non-thermal LOFAR 144~MHz intensities, VLA $L$-band 1.57~GHz intensities, VLA $S$-band 3.0~GHz intensities, and the non-thermal spectral indices $\alpha^{\rm LOFAR}_{S}$ and $\alpha^{L}_{S}$. The red lines represent the best-fit models. The relevant parameters are listed in Table \ref{tab:CR_Fit}.}
    \label{fig:CR_Fit}
\end{figure*}

\begin{figure}[ht]
    \centering
    \includegraphics[width=1\linewidth]{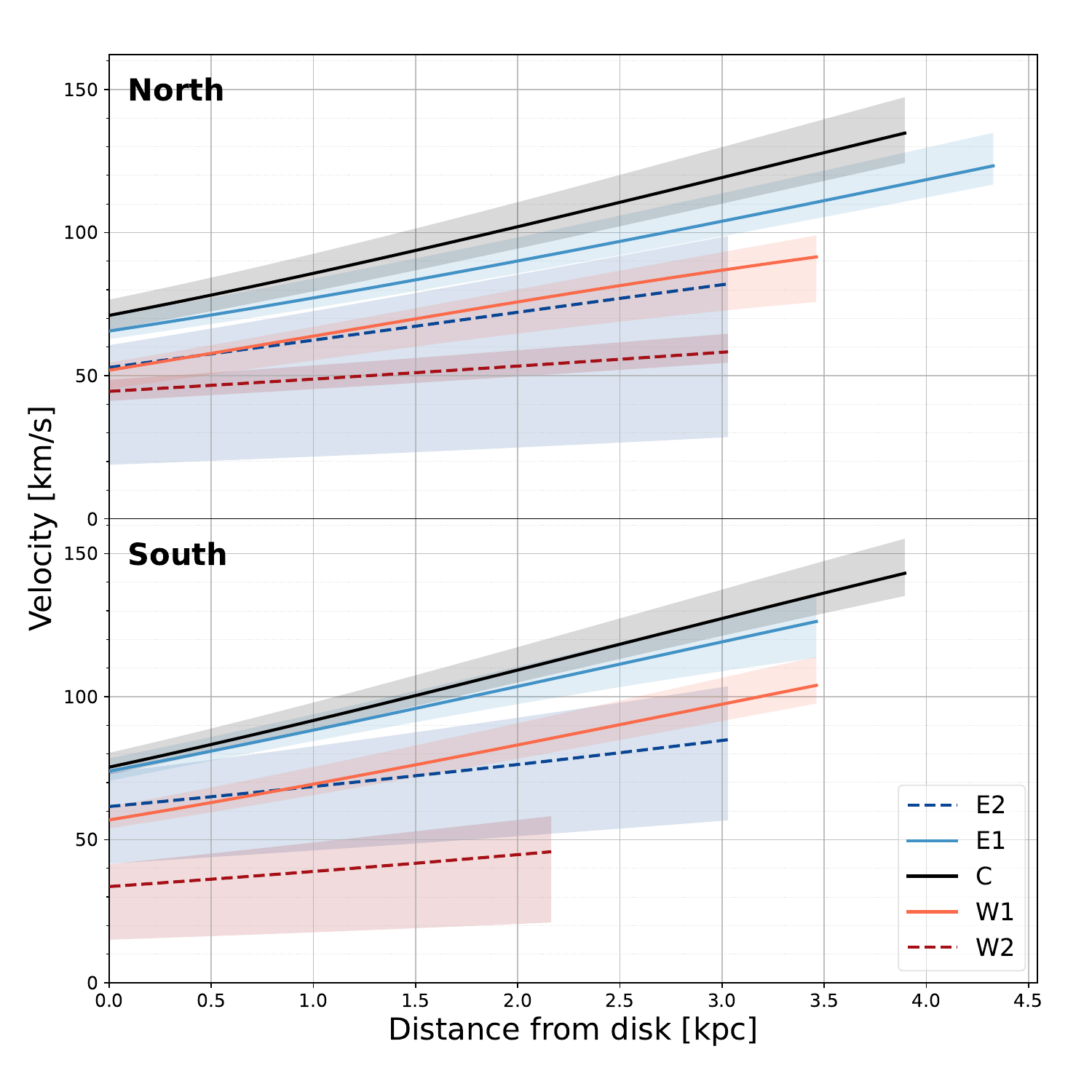}
    \caption{Velocity profiles derived from the advection model. The top and bottom panels display the results for the northern and southern segments, respectively. The solid (E1, C, W1) and dashed (E2, W2) lines show the best-fit solutions, while the shaded areas indicate the $1\sigma$ uncertainties.}
    \label{fig:v_profile}
\end{figure}

\begin{deluxetable*}{lccccccccc}[ht]
\tablecaption{Best-fit Parameters of the CR Transport Models}
\label{tab:CR_Fit}
\tablehead{\colhead{Strip} & \colhead{$\gamma$} & \colhead{$B_0$} & \colhead{$v_c$} & \colhead{$z_0$} & \colhead{$\beta$} & \colhead{$z_c$} & \colhead{$h_{B}$} & \colhead{$D_0$} & \colhead{$\chi^2_{\rm red}$}\\
 & \colhead{} & \colhead{($\mu$G)} & \colhead{(km\,s$^{-1}$)} & \colhead{(kpc)} & \colhead{} & \colhead{(kpc)} & \colhead{(kpc)} & \colhead{($10^{28}$\,cm$^2$\,s$^{-1}$)} & \colhead{}\\
 & \colhead{(fixed)} & \colhead{(fixed)} & \colhead{(adv)} & \colhead{(adv)} & \colhead{(adv)} & \colhead{(adv)} & \colhead{(diff)} & \colhead{(diff)} & \colhead{(adv/diff)}
}
\startdata
E2 N & 2.6 & 4.0 & $164^{+15}_{-15}$ & $10.0^{+16.3}_{-4.3}$ & $2.0^{+2.0}_{-0.3}$ & $18.5^{+26.7}_{-10.3}$ & $3.4^{+0.4}_{-0.4}$ & $2.0^{+0.5}_{-0.3}$ & 0.14/0.25 \\
E2 S & 2.6 & 4.0 & $208^{+8}_{-36}$ & $18.4^{+9.6}_{-10.8}$ & $3.4^{+0.6}_{-2.0}$ & $13.0^{+27.3}_{-3.7}$ & $5.1^{+0.6}_{-0.6}$ & $2.6^{+0.5}_{-0.3}$ & 0.16/0.19 \\
E1 N & 2.6 & 4.6 & $174^{+2}_{-5}$ & $10.3^{+0.5}_{-1.0}$ & $2.4^{+0.1}_{-0.3}$ & $7.9^{+0.9}_{-1.1}$ & $3.3^{+0.1}_{-0.1}$ & $3.4^{+0.3}_{-0.3}$ & 0.71/1.2 \\
E1 S & 2.6 & 4.6 & $170^{+5}_{-9}$ & $8.5^{+0.7}_{-1.3}$ & $2.2^{+0.3}_{-0.3}$ & $6.5^{+24.2}_{-1.0}$ & $3.1^{+0.1}_{-0.1}$ & $3.5^{+0.3}_{-0.3}$ & 0.58/1.1 \\
C N & 2.5 & 4.8 & $166^{+3}_{-3}$ & $8.1^{+0.6}_{-0.5}$ & $2.3^{+0.1}_{-0.1}$ & $5.8^{+0.9}_{-0.8}$ & $2.9^{+0.2}_{-0.1}$ & $3.2^{+0.3}_{-0.4}$ & 1.4/1.6 \\
C S & 2.5 & 4.8 & $164^{+5}_{-7}$ & $7.4^{+0.4}_{-0.8}$ & $2.2^{+0.2}_{-0.3}$ & $5.2^{+19.6}_{-0.7}$ & $2.7^{+0.1}_{-0.1}$ & $4.1^{+0.4}_{-0.3}$ & 1.6/2.9 \\
W1 N & 2.5 & 4.7 & $148^{+6}_{-2}$ & $7.2^{+1.4}_{-0.5}$ & $2.1^{+0.2}_{-0.1}$ & $32.9^{+5.5}_{-24.5}$ & $2.8^{+0.1}_{-0.1}$ & $2.0^{+0.2}_{-0.1}$ & 0.76/2.6 \\
W1 S & 2.5 & 4.7 & $155^{+2}_{-2}$ & $8.2^{+0.4}_{-0.6}$ & $2.3^{+0.1}_{-0.1}$ & $7.5^{+1.0}_{-1.2}$ & $2.9^{+0.1}_{-0.1}$ & $2.0^{+0.2}_{-0.2}$ & 0.95/0.87 \\
W2 N & 2.5 & 3.8 & $220^{+10}_{-53}$ & $32.5^{+12.4}_{-26.3}$ & $3.3^{+0.2}_{-2.1}$ & $24.2^{+28.1}_{-4.8}$ & $7.0^{+1.0}_{-0.8}$ & $1.2^{+0.1}_{-0.1}$ & 0.22/0.49 \\
W2 S & 2.5 & 3.8 & $174^{+17}_{-32}$ & $17.5^{+12.1}_{-12.5}$ & $2.6^{+1.1}_{-1.2}$ & $17.6^{+36.6}_{-4.2}$ & $4.1^{+1.3}_{-0.8}$ & $0.8^{+0.2}_{-0.1}$ & 0.13/0.12 \\
\enddata
\tablecomments{$\gamma$, power-law index of the CR injection spectrum; $B_{0}$, mid-plane magnetic field strength; $v_{c}$, wind speed at the critical point; $z_{0}$, scale height of the flux tube (see Eq. \ref{eq:flux_area}); $\beta$, power-law index of the flux tube; $z_c$, critical height; $h_{B}$, exponential magnetic field scale height; $D_{0}$, energy-independent diffusion coefficient; $\chi^{2}_{\rm red}$, reduced chi-square computed as the arithmetic mean of the fits to the three intensity profiles and two spectral index profiles.}
\end{deluxetable*}

\section{Discussion}\label{sec:discussion}

\subsection{Absence of a Prominent, Extended Radio Halo}

Among all CHANG-ES galaxies, NGC~4565 stands out for its lack of a vertically extended radio halo. Here we discuss several factors that likely contribute to this absence.

The SFR surface density of NGC~4565 was derived by \citet{Vargas19} to be $\Sigma_{\rm SFR} = (0.94\pm0.09)\times10^{-3}\,M_{\odot}\,{\rm yr}^{-1}\,{\rm kpc}^{-2}$. This value is about an order of magnitude lower than in galaxies known to host prominent, extended radio haloes such as NGC~4666 or NGC~5775 \citep[e.g.,][]{Li08,N4666, Heald22}. In the same study \citep[Figure 6 in][]{Vargas19}, a weak positive correlation was found between $\Sigma_{\rm SFR}$ and the halo scale heights (normalized by the radio diameter) in both $C$- and $L$-bands, a trend later confirmed in the $S$-band by \citet{Heesen25}. This correlation is physically reasonable since star formation is the dominant source of cosmic-ray injection. A higher $\Sigma_{\rm SFR}$ implies more supernovae, which drive stronger shocks capable of accelerating cosmic rays \citep{Bell78}. Conversely, a low $\Sigma_{\rm SFR}$ results in fewer supernova explosions and thus a lower injection rate of cosmic rays. It also leads to weaker magnetic fields \citep{Schleicher13, Heesen14}, which directly affect the synchrotron emissivity dominating in the radio band. Indeed, the globally averaged equipartition magnetic field strength in NGC~4565 is relatively weak, $\lesssim5\,\mu{\rm G}$, consistent with this picture.

Another important factor is the galaxy’s mass. More massive galaxies possess deeper gravitational potentials that hinder baryonic outflows from rising above the disk (e.g., \citealt{Li13b}). \citet{Krause18} found an anti-correlation between the normalized $L$-band scale height and the mass surface density, indicating that galaxies with higher mass density tend to have thinner radio haloes. \citet{Schmidt19} extended this analysis by including NGC~891 and NGC~4565, and found that NGC~4565 lies well below the expected correlation, similar to UGC~10288. Both galaxies share the characteristic of relatively weak magnetic fields, suggesting that gravitational confinement and magnetic field strength together play key roles in shaping the halo morphology.

Taken together, the absence of a prominent radio halo in NGC~4565 is likely the result of the combined effects of a low star-formation surface density, a weak magnetic field, and a deep gravitational potential. The relatively inefficient cosmic-ray transport inferred from our modeling (see Section~\ref{sec:CR_model}) further supports this interpretation. Expanding the sample of similar edge-on systems in future works will help to clarify the interplay between these factors in governing halo formation.

\subsection{Magnetic field geometry -- disk-dominated vs local X-shape} \label{sec:Mag_geometry}

The magnetic field in NGC~4565 is largely dominated by disk-parallel components, consistent with its exceptionally thin radio halo. Nevertheless, a localized X-shaped feature is apparent in the northwestern region (quadrants~I and~II in Figure~\ref{fig:PA0_distribution}). The magnetic orientations in this area deviate by approximately $25^{\circ}$ to $30^{\circ}$ from the disk plane in quadrant~I and by $\sim10^{\circ}$ in quadrant~II, forming a pattern reminiscent of the large-scale X-shaped magnetic fields commonly observed in other edge-on galaxies \citep[e.g.,][]{Mora-Partiarroyo19b, Stein20, Stein25}.

The overall disk-dominated configuration can be understood as a natural consequence of the low star-formation activity in NGC~4565. \citet{Stein25} showed that galaxies classified as disk-dominated typically exhibit lower SFR compared to those with prominent X-shaped halos, even though both categories follow the well-established radio–SFR correlation \citep{Condon92, Bell03, Murphy11, Li16, Heesen22, Heesen24b}. Moreover, for the X-shaped galaxies, a higher $\Sigma_{\rm SFR}$ correlates with a larger opening angle of the X-structure, suggesting that stronger star formation leads to more energetic feedback and winds \citep{Heesen18}, which in turn opens the magnetic field lines more widely \citep{Henriksen21}. In this context, the predominantly parallel field of NGC~4565 aligns with its weak star formation and lack of a strong outflow.

The localized X-shaped feature in quadrant~I may instead arise from external influences. \citet{Sancisi08} reported that the \ion{H}{1} emission of NGC~4565 in the velocity range $1250$–$1290\,{\rm km\,s^{-1}}$ bends toward the companion IC~3571, which has a comparable systemic velocity of $1230\,{\rm km\,s^{-1}}$, indicating a tidal disturbance of the outer disk.
The left panel of Figure~\ref{fig:B_HI_X} shows a spatial coincidence between this bending and the magnetic field orientation, suggesting that tidal interaction could locally lift the gas and drag the magnetic field lines into the halo. In quadrant~II, no clear external perturbation is evident, yet the \ion{H}{1} emission shows a subtle curvature along the magnetic field lines (Figure~\ref{fig:B_HI_X}, right). Notably, the soft X-ray emission ($0.5$–$1.5\,{\rm keV}$) extends up to $\sim10\,{\rm kpc}$ above the disk in the same direction \citep{Li13a}. A similar alignment between X-ray outflows and polarized radio structures has been observed in many CHANG-ES galaxies (e.g., NGC~4631 \citealt{Irwin12b}; NGC~3079 \citealt{Li24a}; NGC~3556 \citealt{Li24b}; NGC~4217 \citealt{Stein20}; NGC~5775 \citealt{Heald22}). This may reflect an internally driven, possibly past supernova event that generated a local outflow carrying both hot and neutral gas, with the magnetic fields aligned along the flow direction. 

\begin{deluxetable}{lcc}[t]
\tablecaption{Best-fitting spectral parameters for the diffuse X-ray emission, modeled with \texttt{TBabs} $\times$ \texttt{mxabs} $\times$ (\texttt{apec} + \texttt{powerlaw}).\label{tab:xray_fit_main}}
\tablehead{
\colhead{Parameter} & \colhead{Disk} & \colhead{Corona}
}
\startdata
\multicolumn{3}{c}{\textbf{Absorption}} \\
$N_{\rm H,Gal}$ ($10^{22}\,\mathrm{cm^{-2}}$)\tablenotemark{a} & 0.011 (fixed) & 0.011 (fixed) \\
$N_{\rm H,int}$ ($10^{22}\,\mathrm{cm^{-2}}$) & $2.2^{+1.5}_{-1.2}$ & -- \\
\hline
\multicolumn{3}{c}{\textbf{Thermal Component (apec)}} \\
$kT$ (keV) & $0.27^{+0.03}_{-0.02}$ & 0.27 (fixed)\\
$\rm{norm_{APEC}}$\tablenotemark{b} ($10^{-5}$) & $16.2^{+11.4}_{-8.8}$ & $0.2^{+0.3}_{-0.2}$ \\
\hline
\multicolumn{3}{c}{\textbf{Power-law Component}} \\
Photon Index $\Gamma$ & 1.4 (fixed) & 1.4 (fixed) \\
$\rm{norm_{po}}$\tablenotemark{c} ($10^{-5}$) & $2.9^{+0.6}_{-0.5}$ & $2.5^{+0.5}_{-0.6}$ \\
\hline
C-stat/d.o.f & 97.2/128 & 47.8/53\\
\enddata

\tablenotetext{a}{Galactic foreground $N_{\rm H}$ is adopted from the \textit{HI4PI} all-sky H\,{\sc i} survey \citep{HI4PI2016}.}
\tablenotetext{b}{The XSPEC \texttt{apec} normalization is a dimensionless quantity defined as $\mathrm{norm_{apec}}=10^{-14}[4\pi D_A^2(1+z)^2]^{-1}\int n_{\rm e}n_{\rm H}\,dV$.}
\tablenotetext{c}{The power-law normalization is defined at 1 keV in units of photons ${\rm keV}^{-1} {\rm\,cm}^{-2} {\rm\,s}^{-1}$.}

\end{deluxetable}

We further examine this scenario by comparing the thermal and magnetic pressures in quadrant~II.
The thermal pressure of the hot gas was estimated from the X-ray spectrum.
We extracted the diffuse X-ray spectrum from the polygon region defined by the outermost contour in Figure~8 (corresponding to a surface-brightness level of $1.41\times 10^{-10}{\rm\,photons\,cm^{-2}\,s^{-1}\,arcsec^{-2}}$). Before fitting, we adaptively group each spectrum to achieve a ${\rm S/N} > 1.1$ per bin. The spectrum was fitted with the model \texttt{TBabs} $\times$ \texttt{mxabs} $\times$ (\texttt{apec} + \texttt{powerlaw}), where \texttt{TBabs} accounts for the Galactic foreground absorption (fixed $N_H$ from \citealt{HI4PI2016}), \texttt{mxabs} represents the absorption by the cold interstellar medium of the host galaxy \citep{Luan2025}, \texttt{apec} describes the emission from a collisionally ionized thermal plasma \citep{Smith2001}, and the power-law component accounts for unresolved point sources. The best-fitting spectral parameters, including the temperatures, normalizations, and absorption column densities, are summarized in Table~\ref{tab:xray_fit_main}. Using the best-fit thermal component, we estimated the thermal pressure by approximating the emitting volume as the product of the polygon area and the projected major-axis length of the disk (311\arcsec), and adopting a disk-like geometry with a filling factor of 0.7. 
Under these assumptions, the resulting thermal pressure in the halo region is $P_X = (0.2 \pm 0.1)\,\mathrm{eV\,cm^{-3}}$.

To estimate the magnetic pressure in the same region, we consider that most of this area lies far above the galactic disk (see Figure~\ref{fig:Bmap}). As a reasonable approximation, we adopt the best-fit parameters for the flux-tube advection model of the W1~N strips listed in Table~\ref{tab:CR_Fit}. Taking the height range of the soft X-ray emitting halo to be 4.0 to $8.0\,{\rm kpc}$, this model yields a magnetic field strength that decreases from $2.2$ to $1.4\,\mu{\rm G}$. The resulting magnetic pressure $P_{B} = B^{2}/(2\mu_{0})$ therefore drops from $0.12$ to $0.05\,{\rm eV\,cm^{-3}}$.
We note that this estimate relies on the equipartition assumption, which may not hold in the halos of edge-on galaxies. Recent work by \citet{Chiu25} suggests that equipartition often overestimates the magnetic field strength in such environments. Consequently, the magnetic pressure computed above should be regarded as an upper limit. Even with this conservative estimate, the thermal pressure exceeds the magnetic pressure by a factor of $\sim 2$--$4$. This indicates that the hot gas has sufficient energy density to overcome magnetic confinement and may be capable of advecting magnetic field lines upward, producing the observed extraplanar magnetic structures. The detection of enhanced H$\alpha$ emission in the same region further suggests elevated recent star-formation activity\citep{Mart23}, which could supply the energy necessary to drive the soft X-ray emission and to lift the magnetic field into the halo.

Given the ubiquity of X-shaped magnetic fields observed in many edge-on galaxies, this morphology likely reflects a more general paradigm for galactic magnetic field structures. \citet{Henriksen21} proposed that such configurations can be understood as the two-dimensional projection of rising, spiraling magnetic fields generated by the combined effects of galactic winds and halo lag. For a quiescent system like NGC~4565, where strong outflows are absent or weak, the halo lag mechanism alone may still produce an apparent X-shaped geometry.  Moreover, the model predicts that when halo lag dominates over wind-driven effects, the resulting X-field exhibits a smaller opening angle relative to the disk, which is consistent with the localized feature we observe in NGC~4565.

The spatial correlation between the local RM extrema and the inner radio ring, as highlighted in Section \ref{sec:mag_field}, suggests a coherent magnetic structure coupled with the galactic disk. This feature likely originates from inner spiral arms. Due to the high inclination of NGC~4565, the projected geometry of these spiral arms mimics a ring structure \citep{Schmidt16}. In such an edge-on viewing geometry, the line-of-sight component of an axisymmetric spiral magnetic field is naturally maximized at the tangent points of the arms, which correspond to the edges of the projected ring, resulting in the observed RM peaks on the SE and NW sides \citep{Beck15}. This configuration resembles the magnetic structure found in the central region of NGC~1097, where the magnetic fields trace the spiral pattern within the circumnuclear ring \citep{NGC1097}.

In summary, the magnetic field morphology of NGC~4565 is primarily disk-dominated but shows local coupling with multiphase gas structures. Whether these deviations arise from external tidal interactions or from internally driven outflows and halo lag, the correlation between magnetic geometry and gas morphology indicates that the magnetic field is dynamically linked to the surrounding ISM and CGM. Deeper, high-resolution observations across multiple wavelengths will be essential to constrain the origin and evolution of such localized X-shaped structures.

\begin{figure*}[th!]
    \centering
    \includegraphics[width=\linewidth]{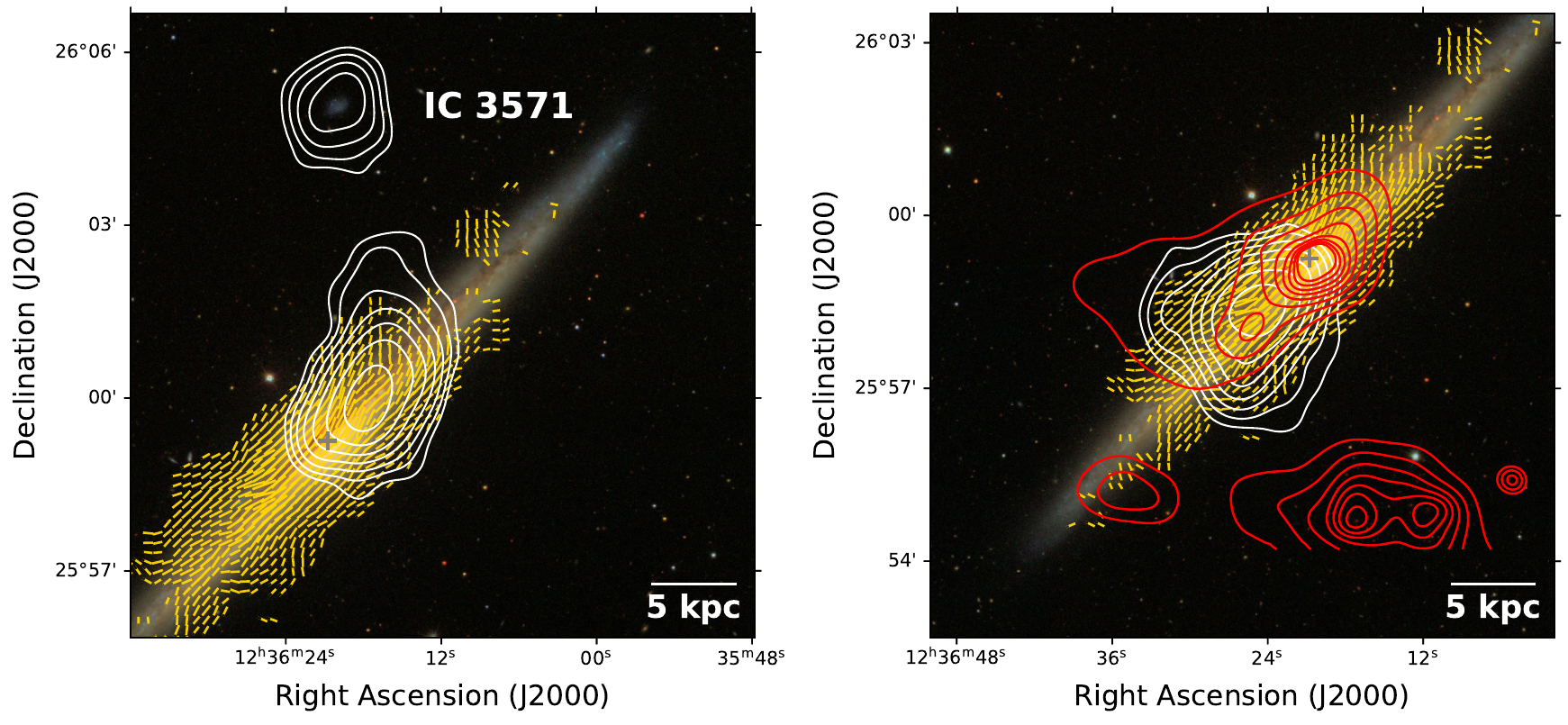}
    \caption{Comparison of magnetic field orientations with multi-phase gas structures in NGC~4565. The \textit{left panel} shows the \ion{H}{1} channel map at $1267\,{\rm km\,s^{-1}}$ \citep{HImap}, displayed as contours overlaid on the SDSS optical image. Contours start at $1\,{\rm mJy\,beam^{-1}}$ and increase by factors of two. The \textit{right panel} presents the soft X-ray emission ($0.5$--$1.5$~keV; red contours), together with the \ion{H}{1} emission at $1164\,{\rm km\,s^{-1}}$ \citep[white contours;][]{HImap} on the same optical background. We note that the distinct X-ray enhancement visible in the bottom-right corner is a background galaxy cluster at a redshift of $\sim 0.62$ \citep{Koulouridis21}. In both panels, magnetic field orientations are shown as yellow lines, and the galactic center is indicated by a cross.}
    \label{fig:B_HI_X}
\end{figure*}

\subsection{The Complex Nature of CR Transport}

Our analysis indicates that CR transport in NGC~4565 is better described by a flux-tube advection model, although signs of complexity remain. While both advection and diffusion models yield statistically acceptable fits to the intensity profiles, with $\chi^{2}_{\rm red}$ close to or below 1, the advection model consistently provides a superior global fit, particularly in reproducing the spectral steepening observed in the spectral index profiles. This finding contrasts with \citet{Heesen19}, who analyzed the older LOFAR data in conjunction with the VLA $L$-band data and concluded that the diffusion model better describes the LOFAR data, whereas the advection provides a superior fit to the $L$-band data. Similarly, \citet{Schmidt19} found the dominant transport mechanism in NGC~4565 to be ambiguous, suggesting that both mechanisms make a significant contribution in the halo.

Furthermore, we notice a radial trend in the best-fit transport parameters. The central strips tend to exhibit slightly higher transport velocities and diffusion coefficients compared to the outer regions (e.g., E2, W2). This behavior is consistent with the higher star formation of the central ring and the activity of the Seyfert nucleus \citep{Muller11} in NGC~4565. These energetic processes provide the necessary power, through supernova explosions, stellar winds, and nuclear outflows, to drive more efficient cosmic ray transport in the central regions.

Despite the overall preference for advection, the comparison at 144~MHz suggests that diffusion may still play a role for the older, lower-energy CRe population. These electrons have had sufficient time to propagate away from their acceleration sites and diffuse into the extended halo. In contrast, the high-frequency 1.57~GHz and 3.0~GHz emission arises from younger, more energetic CRe confined closer to the disk, where their escape is likely governed by a weak, advective outflow or ``breeze'' driven by recent star formation activity.

To test the self-consistency of this interpretation, we compare the characteristic transport time scales with the synchrotron lifetime. The synchrotron lifetime of CRe at 3~GHz in the halo, where the magnetic field is $B \approx 3.5\,\mu{\rm G}$, is $t_{\rm syn} \approx 93\,{\rm Myr}$, following \citet{Krause18}:
\begin{equation}
t_{\rm syn}[{\rm yr}] = 1.06\times10^{9}
\left(\frac{\nu}{\rm GHz}\right)^{-1/2}
\left(\frac{B}{\mu{\rm G}}\right)^{-3/2}.
\end{equation}
For the diffusion model, the diffusion timescale can be estimated as $t_{\rm diff} = l^2 / (4 D_0)$. Adopting the observed vertical extent of the $S$-band emission, $l\approx4.2\,{\rm kpc}$, and the average diffusion coefficient $D_0 \approx 2.5 \times 10^{28}\,{\rm cm^2\,s^{-1}}$, we find a characteristic diffusion time of $t_{\rm diff}\approx53\,{\rm Myr}$. For the advection model, we utilize the average best-fit parameters from all 10 strips. This yields a velocity profile with an initial velocity of $66\,{\rm km\,s^{-1}}$ (consistent with the mean value in Section \ref{sec:CR_model}) and reaching $135\,{\rm km\,s^{-1}}$ at a height of $4.2\,{\rm kpc}$. Integrating this velocity profile results in an advection timescale of $t_{\rm adv}\approx 45\,{\rm Myr}$. Both transport timescales ($t_{\rm diff}$ and $t_{\rm adv}$) are shorter than the synchrotron lifetime ($t_{\rm syn}$). This confirms that both mechanisms are physically capable of transporting cosmic rays to the observed halo height before they lose their energy to radiation.

Overall, these results point to a transport scenario best described by a slow, weakly advective motion. We note, however, that the faintness of the radio halo in NGC~4565 limits the number of independent data points available at high altitudes. Consequently, the current fits cannot definitively distinguish between the advection and diffusion models, particularly for the outer strips. The unprecedented sensitivity and high angular resolution of next-generation interferometers such as the Square Kilometre Array (SKA) will be essential for tracing the halo emission to larger vertical distances and thus better constraining the CR transport processes in such quiescent, disk-dominated systems.

\section{Summary and Conclusions}\label{sec:summary}

In this paper, we present VLA $S$-band (2--4 GHz) radio continuum observations of the CHANG-ES galaxy NGC~4565. Our analysis combines these new data with archival LOFAR, \ion{H}{1}, and soft X-ray observations to provide a comprehensive, multi-wavelength view of the galaxy. In specific, we investigate the magnetic field structure using RM synthesis, characterize the vertical extent of the radio emission, and model the CR transport using the 1D \texttt{SPINNAKER} code. Our results are summarized as follows:

\begin{enumerate}
    \item The $S$-band radio continuum emission of NGC~4565 is vertically compact, lacking a prominent, extended radio halo. Its vertical extent of $\sim 4.2\,{\rm kpc}$ is only about one-sixth of its radial extent. We measure a total integrated flux density at $S$-band of $79.4\pm4.0\,{\rm mJy}$. The absence of a prominent radio halo is likely due to a combination of several factors. The galaxy's low star-formation rate surface density ($\Sigma_{\rm SFR} \approx 0.9\times10^{-3}\,M_\odot\,{\rm yr^{-1}\,kpc^{-2}}$) results in a low CR injection rate and a correspondingly weak magnetic field, while its deep gravitational potential can effectively confine gas and CRs. These conditions inhibit the formation of large-scale outflows necessary to build an extended synchrotron halo. 
    
    \item The vertical intensity profiles are best described by a two-component Gaussian function rather than an exponential model. The resulting average Gaussian scale heights for the disk and halo are $0.82\pm0.04\,{\rm kpc}$ and $3.02\pm0.13\,{\rm kpc}$, respectively.

    \item The magnetic field is predominantly disk-parallel, with relatively weak equipartition strengths of $\sim 4.6\,\mu{\rm G}$ in the disk and $\sim 3.5\,\mu{\rm G}$ in the halo. The RM profile along the major axis reveals an asymmetric distribution with extrema spatially correlating with the inner radio ring, consistent with an axisymmetric spiral magnetic field. Nevertheless, we discover a localized, weak X-shaped pattern in the northeastern region, characterized by magnetic field orientations that deviate more than $10^{\circ}$ from the major axis.
    
    \item We find a compelling spatial coincidence between the vertically oriented magnetic field lines in the northeast and extraplanar extensions seen in both soft X-ray ($0.5\text{--}1.5\,{\rm keV}$) and \ion{H}{1} emission. This morphological alignment suggests a dynamical coupling where outflowing gas drags and reshapes the local magnetic field.

    \item Our 1D CR transport modeling suggests that the flux-tube advection model provides a superior description of the vertical profiles compared to the diffusion model, particularly in reproducing the observed spectral steepening. In this scenario, the transport is characterized by a slow outflow with an initial velocity of $v_0 \approx 60\,{\rm km\,s^{-1}}$, while the diffusion model yields a comparatively small coefficient of $D_0 \approx 2.5\times10^{28}\,{\rm cm^2\,s^{-1}}$. The derived outflow velocities are significantly below the escape velocity, suggesting that CRs and magnetized gas likely remain gravitationally bound, consistent with a galactic fountain scenario.

\end{enumerate}

In conclusion, NGC~4565 serves as a crucial benchmark for studying galactic feedback in a quiescent, low-SFR regime. Its predominantly disk-parallel magnetic field, coupled with inefficient CR transport, highlights the threshold conditions required for the formation of extended radio halos and the launching of galactic winds. Future deep, multi-wavelength observations will be essential for disentangling whether the localized X-shaped field on the northeast side is a nascent galactic wind feature or the result of tidal interactions with a companion.

\begin{acknowledgments}

J.X. and G.L. acknowledge the research grants from the Ministry of Science and Technology of China (National Key Program for Science and Technology Research and Development, No. 2023YFA1608100), the National Natural Science Foundation of China (No. 12273036), and the research grants from the China Manned Space Project (CMS-CSST-2025-A08).
J.T.L. acknowledges the financial support from the science research grants from the China Manned Space Program with grant no. CMS-CSST-2025-A10 and CMS-CSST-2025-A04, and the National Science Foundation of China (NSFC) through grants 12321003 and 12273111. Y.Y. acknowledges support from the National Natural Science Foundation of China (NSFC) through grant No. 12203098. M.S. and R.-J.D. acknowledge funding from the German Science Foundation DFG, within the Collaborative Research Center SFB1491 ``Cosmic Interacting Matters -- From Source to Signal''.

We acknowledge the use of Gemini \citep{Gemini} for checking grammar and improving the readability of this manuscript.

\end{acknowledgments}

\bibliography{citations}{}

@ARTICLE{Becker20,
       author = {{Becker Tjus}, Julia and {Merten}, Lukas},
        title = "{Closing in on the origin of Galactic cosmic rays using multimessenger information}",
      journal = {\physrep},
         year = 2020,
        month = aug,
       volume = {872},
        pages = {1-98},
          doi = {10.1016/j.physrep.2020.05.002},
archivePrefix = {arXiv},
       eprint = {2002.00964},
 primaryClass = {astro-ph.HE},
       adsurl = {https://ui.adsabs.harvard.edu/abs/2020PhR...872....1B}
}

@ARTICLE{Girichidis18,
       author = {{Girichidis}, Philipp and {Naab}, Thorsten and {Hanasz}, Micha{\l} and {Walch}, Stefanie},
        title = "{Cooler and smoother - the impact of cosmic rays on the phase structure of galactic outflows}",
      journal = {\mnras},
         year = 2018,
        month = sep,
       volume = {479},
       number = {3},
        pages = {3042-3067},
          doi = {10.1093/mnras/sty1653},
archivePrefix = {arXiv},
       eprint = {1805.09333},
 primaryClass = {astro-ph.GA},
       adsurl = {https://ui.adsabs.harvard.edu/abs/2018MNRAS.479.3042G}
}

@ARTICLE{Veilleux20,
       author = {{Veilleux}, Sylvain and {Maiolino}, Roberto and {Bolatto}, Alberto D. and {Aalto}, Susanne},
        title = "{Cool outflows in galaxies and their implications}",
      journal = {\aapr},
         year = 2020,
        month = apr,
       volume = {28},
       number = {1},
          eid = {2},
        pages = {2},
          doi = {10.1007/s00159-019-0121-9},
archivePrefix = {arXiv},
       eprint = {2002.07765},
 primaryClass = {astro-ph.GA},
       adsurl = {https://ui.adsabs.harvard.edu/abs/2020A&ARv..28....2V}
}

@ARTICLE{Dobbs08,
       author = {{Dobbs}, C.~L. and {Price}, D.~J.},
        title = "{Magnetic fields and the dynamics of spiral galaxies}",
      journal = {\mnras},
         year = 2008,
        month = jan,
       volume = {383},
       number = {2},
        pages = {497-512},
          doi = {10.1111/j.1365-2966.2007.12591.x},
archivePrefix = {arXiv},
       eprint = {0710.3558},
 primaryClass = {astro-ph},
       adsurl = {https://ui.adsabs.harvard.edu/abs/2008MNRAS.383..497D}
}

@ARTICLE{Krumholz19,
       author = {{Krumholz}, Mark R. and {Federrath}, Christoph},
        title = "{The Role of Magnetic Fields in Setting the Star Formation Rate and the Initial Mass Function}",
      journal = {Frontiers in Astronomy and Space Sciences},
         year = 2019,
        month = feb,
       volume = {6},
          eid = {7},
        pages = {7},
          doi = {10.3389/fspas.2019.00007},
archivePrefix = {arXiv},
       eprint = {1902.02557},
 primaryClass = {astro-ph.GA},
       adsurl = {https://ui.adsabs.harvard.edu/abs/2019FrASS...6....7K}
}

@ARTICLE{Tabatabaei18,
       author = {{Tabatabaei}, F.~S. and {Minguez}, P. and {Prieto}, M.~A. and {Fern{\'a}ndez-Ontiveros}, J.~A.},
        title = "{Discovery of massive star formation quenching by non-thermal effects in the centre of NGC 1097}",
      journal = {Nature Astronomy},
         year = 2018,
        month = nov,
       volume = {2},
        pages = {83-89},
          doi = {10.1038/s41550-017-0298-7},
archivePrefix = {arXiv},
       eprint = {1710.05695},
 primaryClass = {astro-ph.GA},
       adsurl = {https://ui.adsabs.harvard.edu/abs/2018NatAs...2...83T}
}

@ARTICLE{Beck05,
       author = {{Beck}, R. and {Krause}, M.},
        title = "{Revised equipartition and minimum energy formula for magnetic field strength estimates from radio synchrotron observations}",
      journal = {Astronomische Nachrichten},
         year = 2005,
        month = jul,
       volume = {326},
       number = {6},
        pages = {414-427},
          doi = {10.1002/asna.200510366},
archivePrefix = {arXiv},
       eprint = {astro-ph/0507367},
 primaryClass = {astro-ph},
       adsurl = {https://ui.adsabs.harvard.edu/abs/2005AN....326..414B}
}

@ARTICLE{Gilhuly20,
       author = {{Gilhuly}, Colleen and {Hendel}, David and {Merritt}, Allison and {Abraham}, Roberto and {Danieli}, Shany and {Lokhorst}, Deborah and {Liu}, Qing and {van Dokkum}, Pieter and {Conroy}, Charlie and {Greco}, Johnny},
        title = "{The Dragonfly Edge-on Galaxies Survey: Shaping the Outer disk of NGC 4565 via Accretion}",
      journal = {\apj},
         year = 2020,
        month = jul,
       volume = {897},
       number = {2},
          eid = {108},
        pages = {108},
          doi = {10.3847/1538-4357/ab9b25},
archivePrefix = {arXiv},
       eprint = {1910.05358},
 primaryClass = {astro-ph.GA},
       adsurl = {https://ui.adsabs.harvard.edu/abs/2020ApJ...897..108G}
}

@ARTICLE{Heesen19,
       author = {{Heesen}, V. and {Whitler}, L. and {Schmidt}, P. and {Miskolczi}, A. and {Sridhar}, S.~S. and {Horellou}, C. and {Beck}, R. and {G{\"u}rkan}, G. and {Scannapieco}, E. and {Br{\"u}ggen}, M. and {Heald}, G.~H. and {Krause}, M. and {Paladino}, R. and {Nikiel-Wroczy{\'n}ski}, B. and {Wilber}, A. and {Dettmar}, R. -J.},
        title = "{Warped diffusive radio halo around the quiescent spiral edge-on galaxy NGC 4565}",
      journal = {\aap},
         year = 2019,
        month = aug,
       volume = {628},
          eid = {L3},
        pages = {L3},
          doi = {10.1051/0004-6361/201936046},
archivePrefix = {arXiv},
       eprint = {1907.07076},
 primaryClass = {astro-ph.GA},
       adsurl = {https://ui.adsabs.harvard.edu/abs/2019A&A...628L...3H}
}

@ARTICLE{NOD3,
       author = {{M{\"u}ller}, Peter and {Krause}, Marita and {Beck}, Rainer and {Schmidt}, Philip},
        title = "{The NOD3 software package: A graphical user interface-supported reduction package for single-dish radio continuum and polarisation observations}",
      journal = {\aap},
         year = 2017,
        month = oct,
       volume = {606},
          eid = {A41},
        pages = {A41},
          doi = {10.1051/0004-6361/201731257},
archivePrefix = {arXiv},
       eprint = {1707.05573},
 primaryClass = {astro-ph.IM},
       adsurl = {https://ui.adsabs.harvard.edu/abs/2017A&A...606A..41M}
}

@ARTICLE{Dumke95,
       author = {{Dumke}, M. and {Krause}, M. and {Wielebinski}, R. and {Klein}, U.},
        title = "{Polarized radio emission at 2.8cm from a selected sample of edge-on galaxies.}",
      journal = {\aap},
         year = 1995,
        month = oct,
       volume = {302},
        pages = {691},
       adsurl = {https://ui.adsabs.harvard.edu/abs/1995A&A...302..691D}
}

@ARTICLE{Krause18,
       author = {{Krause}, Marita and {Irwin}, Judith and {Wiegert}, Theresa and {Miskolczi}, Arpad and {Damas-Segovia}, Ancor and {Beck}, Rainer and {Li}, Jiang-Tao and {Heald}, George and {M{\"u}ller}, Peter and {Stein}, Yelena and {Rand}, Richard J. and {Heesen}, Volker and {Walterbos}, Rene A.~M. and {Dettmar}, Ralf-J{\"u}rgen and {Vargas}, Carlos J. and {English}, Jayanne and {Murphy}, Eric J.},
        title = "{CHANG-ES. IX. Radio scale heights and scale lengths of a consistent sample of 13 spiral galaxies seen edge-on and their correlations}",
      journal = {\aap},
         year = 2018,
        month = mar,
       volume = {611},
          eid = {A72},
        pages = {A72},
          doi = {10.1051/0004-6361/201731991},
archivePrefix = {arXiv},
       eprint = {1712.03780},
 primaryClass = {astro-ph.GA},
       adsurl = {https://ui.adsabs.harvard.edu/abs/2018A&A...611A..72K}
}

@ARTICLE{Stein23,
       author = {{Stein}, M. and {Heesen}, V. and {Dettmar}, R. -J. and {Stein}, Y. and {Br{\"u}ggen}, M. and {Beck}, R. and {Adebahr}, B. and {Wiegert}, T. and {Vargas}, C.~J. and {Bomans}, D.~J. and {Li}, J. and {English}, J. and {Chy{\.z}y}, K.~T. and {Paladino}, R. and {Tabatabaei}, F.~S. and {Strong}, A.},
        title = "{CHANG-ES. XXVI. Insights into cosmic-ray transport from radio halos in edge-on galaxies}",
      journal = {\aap},
         year = 2023,
        month = feb,
       volume = {670},
          eid = {A158},
        pages = {A158},
          doi = {10.1051/0004-6361/202243906},
archivePrefix = {arXiv},
       eprint = {2210.07709},
 primaryClass = {astro-ph.GA},
       adsurl = {https://ui.adsabs.harvard.edu/abs/2023A&A...670A.158S}
}

@BOOK{circstat,
       author = {{Fisher}, N.~I.},
        title = "{Statistical Analysis of Circular Data}",
         year = 1995,
       adsurl = {https://ui.adsabs.harvard.edu/abs/1995sacd.book.....F}
}

@ARTICLE{Krause20,
       author = {{Krause}, Marita and {Irwin}, Judith and {Schmidt}, Philip and {Stein}, Yelena and {Miskolczi}, Arpad and {Carolina Mora-Partiarroyo}, Silvia and {Wiegert}, Theresa and {Beck}, Rainer and {Stil}, Jeroen M. and {Heald}, George and {Li}, Jiang-Tao and {Damas-Segovia}, Ancor and {Vargas}, Carlos J. and {Rand}, Richard J. and {West}, Jennifer and {Walterbos}, Rene A.~M. and {Dettmar}, Ralf-J{\"u}rgen and {English}, Jayanne and {Woodfinden}, Alex},
        title = "{CHANG-ES. XXII. Coherent magnetic fields in the halos of spiral galaxies}",
      journal = {\aap},
         year = 2020,
        month = jul,
       volume = {639},
          eid = {A112},
        pages = {A112},
          doi = {10.1051/0004-6361/202037780},
archivePrefix = {arXiv},
       eprint = {2004.14383},
 primaryClass = {astro-ph.GA},
       adsurl = {https://ui.adsabs.harvard.edu/abs/2020A&A...639A.112K}
}

@ARTICLE{Stein25,
       author = {{Stein}, M. and {Kleimann}, J. and {Adebahr}, B. and {Dettmar}, R. -J. and {Fichtner}, H. and {English}, J. and {Heesen}, V. and {Kamphuis}, P. and {Irwin}, J. and {Mele}, C. and {Bomans}, D.~J. and {Li}, J. and {Skeggs}, N.~B. and {Wang}, Q.~D. and {Yang}, Y.},
        title = "{CHANG-ES: XXXIV. Magnetic field structure in edge-on galaxies: Characterising large-scale magnetic fields in galactic halos}",
      journal = {\aap},
         year = 2025,
        month = apr,
       volume = {696},
          eid = {A112},
        pages = {A112},
          doi = {10.1051/0004-6361/202452322},
archivePrefix = {arXiv},
       eprint = {2503.05461},
 primaryClass = {astro-ph.GA},
       adsurl = {https://ui.adsabs.harvard.edu/abs/2025A&A...696A.112S}
}

@ARTICLE{Sukumar91,
       author = {{Sukumar}, S. and {Allen}, R.~J.},
        title = "{Polarized Radio Emission from the Edge-on Spiral Galaxies NGC 891 and NGC 4565}",
      journal = {\apj},
         year = 1991,
        month = nov,
       volume = {382},
        pages = {100},
          doi = {10.1086/170697},
       adsurl = {https://ui.adsabs.harvard.edu/abs/1991ApJ...382..100S}
}

@ARTICLE{Sancisi08,
       author = {{Sancisi}, Renzo and {Fraternali}, Filippo and {Oosterloo}, Tom and {van der Hulst}, Thijs},
        title = "{Cold gas accretion in galaxies}",
      journal = {\aapr},
         year = 2008,
        month = jun,
       volume = {15},
       number = {3},
        pages = {189-223},
          doi = {10.1007/s00159-008-0010-0},
archivePrefix = {arXiv},
       eprint = {0803.0109},
 primaryClass = {astro-ph},
       adsurl = {https://ui.adsabs.harvard.edu/abs/2008A&ARv..15..189S}
}

@PHDTHESIS{Schmidt16,
       author = {{Schmidt}, Philip},
        title = "{The radio continuum halos of the edge-on galaxies NGC 891 and NGC 4565}",
       school = {Rheinische Friedrich Wilhelms University of Bonn, Germany},
         year = 2016,
        month = jan,
       adsurl = {https://ui.adsabs.harvard.edu/abs/2016PhDT.......560S}
}

@ARTICLE{Xu25,
       author = {{Xu}, Jianghui and {Yang}, Yang and {Li}, Jiang-Tao and {Liu}, Guilin and {Irwin}, Judith and {Dettmar}, Ralf-J{\"u}rgen and {Stein}, Michael and {Wiegert}, Theresa and {Wang}, Q. Daniel and {English}, Jayanne},
        title = "{CHANG-ES. XXXV. Cosmic Ray Transport and Magnetic Field Structure of NGC 3556 at 3 GHz}",
      journal = {\apj},
         year = 2025,
        month = jan,
       volume = {978},
       number = {1},
          eid = {5},
        pages = {5},
          doi = {10.3847/1538-4357/ad946e},
archivePrefix = {arXiv},
       eprint = {2411.12564},
 primaryClass = {astro-ph.GA},
       adsurl = {https://ui.adsabs.harvard.edu/abs/2025ApJ...978....5X}
}

@ARTICLE{RM-tools26,
       author = {{Van Eck}, Cameron L. and {Purcell}, Cormac R. and {Baidoo}, Lerato and {Thomson}, Alec J.~M. and {Ma}, Yik Ki and {Oberhelman}, Lindsey and {Osinga}, Erik and {Vanderwoude}, Shannon and {West}, Jennifer L. and {Ideguchi}, Shinsuke and {Par{\'e}}, Dylan M. and {Kaczmarek}, Jane F. and {Willis}, Tony and {Akahori}, Takuya and {Anderson}, Craig S. and {Gaensler}, B.~M. and {O'Sullivan}, Shane and {Sun}, Xiaohui and {Amaral}, Ariel D. and {Riseley}, C.~J. and {Stil}, Jeroen and {Zhang}, Xiang},
        title = "{RM-Tools: Software for Analyzing Polarized Radio Spectra}",
      journal = {arXiv e-prints},
         year = 2026,
        month = jan,
          eid = {arXiv:2601.20092},
        pages = {arXiv:2601.20092},
          doi = {10.48550/arXiv.2601.20092},
archivePrefix = {arXiv},
       eprint = {2601.20092},
 primaryClass = {astro-ph.IM},
       adsurl = {https://ui.adsabs.harvard.edu/abs/2026arXiv260120092V}
}

@software{RM-Tools,
       author = {{Purcell}, C.~R. and {Van Eck}, C.~L. and {West}, J. and {Sun}, X.~H. and {Gaensler}, B.~M.},
        title = "{RM-Tools: Rotation measure (RM) synthesis and Stokes QU-fitting}",
 howpublished = {Astrophysics Source Code Library, record ascl:2005.003},
         year = 2020,
        month = may,
          eid = {ascl:2005.003},
       adsurl = {https://ui.adsabs.harvard.edu/abs/2020ascl.soft05003P}
}

@ARTICLE{Burn96,
       author = {{Burn}, B.~J.},
        title = "{On the depolarization of discrete radio sources by Faraday dispersion}",
      journal = {\mnras},
         year = 1966,
        month = jan,
       volume = {133},
        pages = {67},
          doi = {10.1093/mnras/133.1.67},
       adsurl = {https://ui.adsabs.harvard.edu/abs/1966MNRAS.133...67B}
}

@ARTICLE{RMsyn,
       author = {{Brentjens}, M.~A. and {de Bruyn}, A.~G.},
        title = "{Faraday rotation measure synthesis}",
      journal = {\aap},
         year = 2005,
        month = oct,
       volume = {441},
       number = {3},
        pages = {1217-1228},
          doi = {10.1051/0004-6361:20052990},
archivePrefix = {arXiv},
       eprint = {astro-ph/0507349},
 primaryClass = {astro-ph},
       adsurl = {https://ui.adsabs.harvard.edu/abs/2005A&A...441.1217B}
}

@ARTICLE{IMFIT,
       author = {{Erwin}, Peter},
        title = "{IMFIT: A Fast, Flexible New Program for Astronomical Image Fitting}",
      journal = {\apj},
         year = 2015,
        month = feb,
       volume = {799},
       number = {2},
          eid = {226},
        pages = {226},
          doi = {10.1088/0004-637X/799/2/226},
archivePrefix = {arXiv},
       eprint = {1408.1097},
 primaryClass = {astro-ph.IM},
       adsurl = {https://ui.adsabs.harvard.edu/abs/2015ApJ...799..226E}
}

@ARTICLE{Neininger96,
       author = {{Neininger}, N. and {Guelin}, M. and {Garcia-Burillo}, S. and {Zylka}, R. and {Wielebinski}, R.},
        title = "{Cold dust and molecular line emission in NGC4565.}",
      journal = {\aap},
         year = 1996,
        month = jun,
       volume = {310},
        pages = {725-736},
       adsurl = {https://ui.adsabs.harvard.edu/abs/1996A&A...310..725N}
}

@ARTICLE{Kormendy10,
       author = {{Kormendy}, John and {Barentine}, John C.},
        title = "{Detection of a Pseudobulge Hidden Inside the ``Box-shaped Bulge'' of NGC 4565}",
      journal = {\apjl},
         year = 2010,
        month = jun,
       volume = {715},
       number = {2},
        pages = {L176-L179},
          doi = {10.1088/2041-8205/715/2/L176},
archivePrefix = {arXiv},
       eprint = {1005.1647},
 primaryClass = {astro-ph.GA},
       adsurl = {https://ui.adsabs.harvard.edu/abs/2010ApJ...715L.176K}
}

@ARTICLE{Laine10,
       author = {{Laine}, Seppo and {Appleton}, Philip N. and {Gottesman}, Stephen T. and {Ashby}, Matthew L.~N. and {Garland}, Catherine A.},
        title = "{Warm Molecular Hydrogen Emission in Normal Edge-on Galaxies NGC 4565 and NGC 5907}",
      journal = {\aj},
         year = 2010,
        month = sep,
       volume = {140},
       number = {3},
        pages = {753-769},
          doi = {10.1088/0004-6256/140/3/753},
archivePrefix = {arXiv},
       eprint = {1007.4194},
 primaryClass = {astro-ph.CO},
       adsurl = {https://ui.adsabs.harvard.edu/abs/2010AJ....140..753L}
}

@ARTICLE{Yim14,
       author = {{Yim}, Kijeong and {Wong}, Tony and {Xue}, Rui and {Rand}, Richard J. and {Rosolowsky}, Erik and {van der Hulst}, J.~M. and {Benjamin}, Robert and {Murphy}, Eric J.},
        title = "{The Interstellar Medium and Star Formation in Edge-On Galaxies. II. NGC 4157, 4565, and 5907}",
      journal = {\aj},
         year = 2014,
        month = dec,
       volume = {148},
       number = {6},
          eid = {127},
        pages = {127},
          doi = {10.1088/0004-6256/148/6/127},
archivePrefix = {arXiv},
       eprint = {1408.5905},
 primaryClass = {astro-ph.GA},
       adsurl = {https://ui.adsabs.harvard.edu/abs/2014AJ....148..127Y}
}

@ARTICLE{Schmidt19,
       author = {{Schmidt}, Philip and {Krause}, Marita and {Heesen}, Volker and {Basu}, Aritra and {Beck}, Rainer and {Wiegert}, Theresa and {Irwin}, Judith A. and {Heald}, George and {Rand}, Richard J. and {Li}, Jiang-Tao and {Murphy}, Eric J.},
        title = "{CHANG-ES. XVI. An in-depth view of the cosmic-ray transport in the edge-on spiral galaxies NGC 891 and NGC 4565}",
      journal = {\aap},
         year = 2019,
        month = dec,
       volume = {632},
          eid = {A12},
        pages = {A12},
          doi = {10.1051/0004-6361/201834995},
archivePrefix = {arXiv},
       eprint = {1907.03789},
 primaryClass = {astro-ph.GA},
       adsurl = {https://ui.adsabs.harvard.edu/abs/2019A&A...632A..12S}
}

@ARTICLE{Lambrides19,
       author = {{Lambrides}, Erini L. and {Petric}, Andreea O. and {Tchernyshyov}, Kirill and {Zakamska}, Nadia L. and {Watts}, Duncan J.},
        title = "{Mid-infrared spectroscopic evidence for AGN heating warm molecular gas}",
      journal = {\mnras},
         year = 2019,
        month = aug,
       volume = {487},
       number = {2},
        pages = {1823-1843},
          doi = {10.1093/mnras/stz1316},
archivePrefix = {arXiv},
       eprint = {1808.02035},
 primaryClass = {astro-ph.GA},
       adsurl = {https://ui.adsabs.harvard.edu/abs/2019MNRAS.487.1823L}
}

@ARTICLE{Irwin19a,
       author = {{Irwin}, Judith and {Wiegert}, Theresa and {Merritt}, Alison and {We{\.z}gowiec}, Marek and {Hunt}, Lucas and {Woodfinden}, Alex and {Stein}, Yelena and {Damas-Segovia}, Ancor and {Li}, Jiangtao and {Wang}, Q. Daniel and {Johnson}, Megan and {Krause}, Marita and {Dettmar}, Ralf-J{\"u}rgen and {Im}, Jisung and {Schmidt}, Philip and {Miskolczi}, Arpad and {Braun}, Timothy T. and {Saikia}, D.~J. and {English}, Jayanne and {Richardson}, Mark},
        title = "{CHANG-ES. XX. High-resolution Radio Continuum Images of Edge-on Galaxies and Their AGNs: Data Release 3}",
      journal = {\aj},
         year = 2019,
        month = jul,
       volume = {158},
       number = {1},
          eid = {21},
        pages = {21},
          doi = {10.3847/1538-3881/ab25f6},
archivePrefix = {arXiv},
       eprint = {1905.05160},
 primaryClass = {astro-ph.GA},
       adsurl = {https://ui.adsabs.harvard.edu/abs/2019AJ....158...21I}
}

@ARTICLE{Bhatnagar08,
       author = {{Bhatnagar}, S. and {Cornwell}, T.~J. and {Golap}, K. and {Uson}, J.~M.},
        title = "{Correcting direction-dependent gains in the deconvolution of radio interferometric images}",
      journal = {\aap},
         year = 2008,
        month = aug,
       volume = {487},
       number = {1},
        pages = {419-429},
          doi = {10.1051/0004-6361:20079284},
archivePrefix = {arXiv},
       eprint = {0805.0834},
 primaryClass = {astro-ph},
       adsurl = {https://ui.adsabs.harvard.edu/abs/2008A&A...487..419B}
}

@ARTICLE{mtmfs,
       author = {{Rau}, U. and {Cornwell}, T.~J.},
        title = "{A multi-scale multi-frequency deconvolution algorithm for synthesis imaging in radio interferometry}",
      journal = {\aap},
         year = 2011,
        month = aug,
       volume = {532},
          eid = {A71},
        pages = {A71},
          doi = {10.1051/0004-6361/201117104},
archivePrefix = {arXiv},
       eprint = {1106.2745},
 primaryClass = {astro-ph.IM},
       adsurl = {https://ui.adsabs.harvard.edu/abs/2011A&A...532A..71R}
}

@ARTICLE{CASA22,
       author = {{CASA Team} and {Bean}, Ben and {Bhatnagar}, Sanjay and {Castro}, Sandra and {Donovan Meyer}, Jennifer and {Emonts}, Bjorn and {Garcia}, Enrique and {Garwood}, Robert and {Golap}, Kumar and {Gonzalez Villalba}, Justo and {Harris}, Pamela and {Hayashi}, Yohei and {Hoskins}, Josh and {Hsieh}, Mingyu and {Jagannathan}, Preshanth and {Kawasaki}, Wataru and {Keimpema}, Aard and {Kettenis}, Mark and {Lopez}, Jorge and {Marvil}, Joshua and {Masters}, Joseph and {McNichols}, Andrew and {Mehringer}, David and {Miel}, Renaud and {Moellenbrock}, George and {Montesino}, Federico and {Nakazato}, Takeshi and {Ott}, Juergen and {Petry}, Dirk and {Pokorny}, Martin and {Raba}, Ryan and {Rau}, Urvashi and {Schiebel}, Darrell and {Schweighart}, Neal and {Sekhar}, Srikrishna and {Shimada}, Kazuhiko and {Small}, Des and {Steeb}, Jan-Willem and {Sugimoto}, Kanako and {Suoranta}, Ville and {Tsutsumi}, Takahiro and {van Bemmel}, Ilse M. and {Verkouter}, Marjolein and {Wells}, Akeem and {Xiong}, Wei and {Szomoru}, Arpad and {Griffith}, Morgan and {Glendenning}, Brian and {Kern}, Jeff},
        title = "{CASA, the Common Astronomy Software Applications for Radio Astronomy}",
      journal = {\pasp},
         year = 2022,
        month = nov,
       volume = {134},
       number = {1041},
          eid = {114501},
        pages = {114501},
          doi = {10.1088/1538-3873/ac9642},
archivePrefix = {arXiv},
       eprint = {2210.02276},
 primaryClass = {astro-ph.IM},
       adsurl = {https://ui.adsabs.harvard.edu/abs/2022PASP..134k4501C}
}

@ARTICLE{Irwin12,
       author = {{Irwin}, Judith and {Beck}, Rainer and {Benjamin}, R.~A. and {Dettmar}, Ralf-J{\"u}rgen and {English}, Jayanne and {Heald}, George and {Henriksen}, Richard N. and {Johnson}, Megan and {Krause}, Marita and {Li}, Jiang-Tao and {Miskolczi}, Arpad and {Mora}, Silvia Carolina and {Murphy}, E.~J. and {Oosterloo}, Tom and {Porter}, Troy A. and {Rand}, Richard J. and {Saikia}, D.~J. and {Schmidt}, Philip and {Strong}, A.~W. and {Walterbos}, Rene and {Wang}, Q. Daniel and {Wiegert}, Theresa},
        title = "{Continuum Halos in Nearby Galaxies: An EVLA Survey (CHANG-ES). I. Introduction to the Survey}",
      journal = {\aj},
         year = 2012,
        month = aug,
       volume = {144},
       number = {2},
          eid = {43},
        pages = {43},
          doi = {10.1088/0004-6256/144/2/43},
archivePrefix = {arXiv},
       eprint = {1205.5694},
 primaryClass = {astro-ph.CO},
       adsurl = {https://ui.adsabs.harvard.edu/abs/2012AJ....144...43I}
}

@software{colormap,
       author = {{English}, Jayanne and {Richardson}, Mark L.~A. and {Ferrand}, Gilles and {Deg}, Nathan},
        title = "{CosmosCanvas: Useful color maps for different astrophysical properties}",
 howpublished = {Astrophysics Source Code Library, record ascl:2401.005},
         year = 2024,
        month = jan,
          eid = {ascl:2401.005},
       adsurl = {https://ui.adsabs.harvard.edu/abs/2024ascl.soft01005E}
}

@ARTICLE{Wiegert15,
       author = {{Wiegert}, Theresa and {Irwin}, Judith and {Miskolczi}, Arpad and {Schmidt}, Philip and {Mora}, Silvia Carolina and {Damas-Segovia}, Ancor and {Stein}, Yelena and {English}, Jayanne and {Rand}, Richard J. and {Santistevan}, Isaiah and {Walterbos}, Rene and {Krause}, Marita and {Beck}, Rainer and {Dettmar}, Ralf-J{\"u}rgen and {Kepley}, Amanda and {Wezgowiec}, Marek and {Wang}, Q. Daniel and {Heald}, George and {Li}, Jiangtao and {MacGregor}, Stephen and {Johnson}, Megan and {Strong}, A.~W. and {DeSouza}, Amanda and {Porter}, Troy A.},
        title = "{CHANG-ES. IV. Radio Continuum Emission of 35 Edge-on Galaxies Observed with the Karl G. Jansky Very Large Array in D Configuration{\textemdash}Data Release 1}",
      journal = {\aj},
         year = 2015,
        month = sep,
       volume = {150},
       number = {3},
          eid = {81},
        pages = {81},
          doi = {10.1088/0004-6256/150/3/81},
archivePrefix = {arXiv},
       eprint = {1508.05153},
 primaryClass = {astro-ph.GA},
       adsurl = {https://ui.adsabs.harvard.edu/abs/2015AJ....150...81W}
}

@ARTICLE{Heesen25,
       author = {{Heesen}, V. and {Stein}, M. and {Pourjafari}, N. and {Br{\"u}ggen}, M. and {Stil}, J. and {Li}, J. -T. and {Wiegert}, T. and {Irwin}, J. and {Dettmar}, R. -J. and {Porter}, T.~A. and {Stein}, Y.},
        title = "{CHANG-ES: XXXVI. The thin and thick radio discs}",
      journal = {\aap},
         year = 2025,
        month = jul,
       volume = {699},
          eid = {A243},
        pages = {A243},
          doi = {10.1051/0004-6361/202554046},
archivePrefix = {arXiv},
       eprint = {2505.13713},
 primaryClass = {astro-ph.GA},
       adsurl = {https://ui.adsabs.harvard.edu/abs/2025A&A...699A.243H}
}

@ARTICLE{Boulares90,
       author = {{Boulares}, Ahmed and {Cox}, Donald P.},
        title = "{Galactic Hydrostatic Equilibrium with Magnetic Tension and Cosmic-Ray Diffusion}",
      journal = {\apj},
         year = 1990,
        month = dec,
       volume = {365},
        pages = {544},
          doi = {10.1086/169509},
       adsurl = {https://ui.adsabs.harvard.edu/abs/1990ApJ...365..544B}
}

@ARTICLE{Heald22,
       author = {{Heald}, G.~H. and {Heesen}, V. and {Sridhar}, S.~S. and {Beck}, R. and {Bomans}, D.~J. and {Br{\"u}ggen}, M. and {Chy{\.z}y}, K.~T. and {Damas-Segovia}, A. and {Dettmar}, R. -J. and {English}, J. and {Henriksen}, R. and {Ideguchi}, S. and {Irwin}, J. and {Krause}, M. and {Li}, J. -T. and {Murphy}, E.~J. and {Nikiel-Wroczy{\'n}ski}, B. and {Piotrowska}, J. and {Rand}, R.~J. and {Shimwell}, T. and {Stein}, Y. and {Vargas}, C.~J. and {Wang}, Q.~D. and {van Weeren}, R.~J. and {Wiegert}, T.},
        title = "{CHANG-ES XXIII: influence of a galactic wind in NGC 5775}",
      journal = {\mnras},
         year = 2022,
        month = jan,
       volume = {509},
       number = {1},
        pages = {658-684},
          doi = {10.1093/mnras/stab2804},
archivePrefix = {arXiv},
       eprint = {2109.12267},
 primaryClass = {astro-ph.GA},
       adsurl = {https://ui.adsabs.harvard.edu/abs/2022MNRAS.509..658H}
}

@ARTICLE{Mora-Partiarroyo19,
       author = {{Mora-Partiarroyo}, Silvia Carolina and {Krause}, Marita and {Basu}, Aritra and {Beck}, Rainer and {Wiegert}, Theresa and {Irwin}, Judith and {Henriksen}, Richard and {Stein}, Yelena and {Vargas}, Carlos J. and {Heesen}, Volker and {Walterbos}, Ren{\'e} A.~M. and {Rand}, Richard J. and {Heald}, George and {Li}, Jiangtao and {Kamieneski}, Patrick and {English}, Jayanne},
        title = "{CHANG-ES. XIV. Cosmic-ray propagation and magnetic field strengths in the radio halo of NGC 4631}",
      journal = {\aap},
         year = 2019,
        month = dec,
       volume = {632},
          eid = {A10},
        pages = {A10},
          doi = {10.1051/0004-6361/201834571},
archivePrefix = {arXiv},
       eprint = {1910.07588},
 primaryClass = {astro-ph.GA},
       adsurl = {https://ui.adsabs.harvard.edu/abs/2019A&A...632A..10M}
}

@ARTICLE{Beck15,
       author = {{Beck}, Rainer},
        title = "{Magnetic fields in spiral galaxies}",
      journal = {\aapr},
         year = 2015,
        month = dec,
       volume = {24},
          eid = {4},
        pages = {4},
          doi = {10.1007/s00159-015-0084-4},
archivePrefix = {arXiv},
       eprint = {1509.04522},
 primaryClass = {astro-ph.GA},
       adsurl = {https://ui.adsabs.harvard.edu/abs/2015A&ARv..24....4B}
}

@ARTICLE{Han17,
       author = {{Han}, J.~L.},
        title = "{Observing Interstellar and Intergalactic Magnetic Fields}",
      journal = {\araa},
         year = 2017,
        month = aug,
       volume = {55},
       number = {1},
        pages = {111-157},
          doi = {10.1146/annurev-astro-091916-055221},
       adsurl = {https://ui.adsabs.harvard.edu/abs/2017ARA&A..55..111H}
}

@ARTICLE{Mora-Partiarroyo19b,
       author = {{Mora-Partiarroyo}, Silvia Carolina and {Krause}, Marita and {Basu}, Aritra and {Beck}, Rainer and {Wiegert}, Theresa and {Irwin}, Judith and {Henriksen}, Richard and {Stein}, Yelena and {Vargas}, Carlos J. and {Heesen}, Volker and {Walterbos}, Ren{\'e} A.~M. and {Rand}, Richard J. and {Heald}, George and {Li}, Jiangtao and {Kamieneski}, Patrick and {English}, Jayanne},
        title = "{CHANG-ES. XV. Large-scale magnetic field reversals in the radio halo of NGC 4631}",
      journal = {\aap},
         year = 2019,
        month = dec,
       volume = {632},
          eid = {A11},
        pages = {A11},
          doi = {10.1051/0004-6361/201935961},
archivePrefix = {arXiv},
       eprint = {1910.07590},
 primaryClass = {astro-ph.GA},
       adsurl = {https://ui.adsabs.harvard.edu/abs/2019A&A...632A..11M}
}

@ARTICLE{Stein20,
       author = {{Stein}, Y. and {Dettmar}, R. -J. and {Beck}, R. and {Irwin}, J. and {Wiegert}, T. and {Miskolczi}, A. and {Wang}, Q.~D. and {English}, J. and {Henriksen}, R. and {Radica}, M. and {Li}, J. -T.},
        title = "{CHANG-ES. XXI. Transport processes and the X-shaped magnetic field of NGC 4217: off-center superbubble structure}",
      journal = {\aap},
         year = 2020,
        month = jul,
       volume = {639},
          eid = {A111},
        pages = {A111},
          doi = {10.1051/0004-6361/202037675},
archivePrefix = {arXiv},
       eprint = {2007.03002},
 primaryClass = {astro-ph.GA},
       adsurl = {https://ui.adsabs.harvard.edu/abs/2020A&A...639A.111S}
}

@ARTICLE{Irwin12b,
       author = {{Irwin}, Judith and {Beck}, Rainer and {Benjamin}, R.~A. and {Dettmar}, Ralf-J{\"u}rgen and {English}, Jayanne and {Heald}, George and {Henriksen}, Richard N. and {Johnson}, Megan and {Krause}, Marita and {Li}, Jiang-Tao and {Miskolczi}, Arpad and {Mora}, Silvia Carolina and {Murphy}, E.~J. and {Oosterloo}, Tom and {Porter}, Troy A. and {Rand}, Richard J. and {Saikia}, D.~J. and {Schmidt}, Philip and {Strong}, A.~W. and {Walterbos}, Rene and {Wang}, Q. Daniel and {Wiegert}, Theresa},
        title = "{Continuum Halos in Nearby Galaxies: An EVLA Survey (CHANG-ES). II. First Results on NGC 4631}",
      journal = {\aj},
         year = 2012,
        month = aug,
       volume = {144},
       number = {2},
          eid = {44},
        pages = {44},
          doi = {10.1088/0004-6256/144/2/44},
archivePrefix = {arXiv},
       eprint = {1205.5771},
 primaryClass = {astro-ph.GA},
       adsurl = {https://ui.adsabs.harvard.edu/abs/2012AJ....144...44I}
}

@ARTICLE{Vargas19,
       author = {{Vargas}, Carlos J. and {Walterbos}, Ren{\'e} A.~M. and {Rand}, Richard J. and {Stil}, Jeroen and {Krause}, Marita and {Li}, Jiang-Tao and {Irwin}, Judith and {Dettmar}, Ralf-J{\"u}rgen},
        title = "{CHANG-ES. XVII. H{\ensuremath{\alpha}} Imaging of Nearby Edge-on Galaxies, New SFRs, and an Extreme Star Formation Region{\textemdash}Data Release 2}",
      journal = {\apj},
         year = 2019,
        month = aug,
       volume = {881},
       number = {1},
          eid = {26},
        pages = {26},
          doi = {10.3847/1538-4357/ab27cb},
archivePrefix = {arXiv},
       eprint = {1906.07763},
 primaryClass = {astro-ph.GA},
       adsurl = {https://ui.adsabs.harvard.edu/abs/2019ApJ...881...26V}
}

@ARTICLE{Heesen16,
       author = {{Heesen}, Volker and {Dettmar}, Ralf-J{\"u}rgen and {Krause}, Marita and {Beck}, Rainer and {Stein}, Yelena},
        title = "{Advective and diffusive cosmic ray transport in galactic haloes}",
      journal = {\mnras},
         year = 2016,
        month = may,
       volume = {458},
       number = {1},
        pages = {332-353},
          doi = {10.1093/mnras/stw360},
archivePrefix = {arXiv},
       eprint = {1602.04085},
 primaryClass = {astro-ph.GA},
       adsurl = {https://ui.adsabs.harvard.edu/abs/2016MNRAS.458..332H}
}

@ARTICLE{Heesen18,
       author = {{Heesen}, V. and {Krause}, M. and {Beck}, R. and {Adebahr}, B. and {Bomans}, D.~J. and {Carretti}, E. and {Dumke}, M. and {Heald}, G. and {Irwin}, J. and {Koribalski}, B.~S. and {Mulcahy}, D.~D. and {Westmeier}, T. and {Dettmar}, R. -J.},
        title = "{Radio haloes in nearby galaxies modelled with 1D cosmic ray transport using SPINNAKER}",
      journal = {\mnras},
         year = 2018,
        month = may,
       volume = {476},
       number = {1},
        pages = {158-183},
          doi = {10.1093/mnras/sty105},
archivePrefix = {arXiv},
       eprint = {1801.05211},
 primaryClass = {astro-ph.GA},
       adsurl = {https://ui.adsabs.harvard.edu/abs/2018MNRAS.476..158H}
}

@ARTICLE{Li13a,
       author = {{Li}, Jiang-Tao and {Wang}, Q. Daniel},
        title = "{Chandra survey of nearby highly inclined disc galaxies - I. X-ray measurements of galactic coronae}",
      journal = {\mnras},
         year = 2013,
        month = jan,
       volume = {428},
       number = {3},
        pages = {2085-2108},
          doi = {10.1093/mnras/sts183},
archivePrefix = {arXiv},
       eprint = {1210.2997},
 primaryClass = {astro-ph.CO},
       adsurl = {https://ui.adsabs.harvard.edu/abs/2013MNRAS.428.2085L}
}

@ARTICLE{Bell78,
       author = {{Bell}, A.~R.},
        title = "{The acceleration of cosmic rays in shock fronts - I.}",
      journal = {\mnras},
         year = 1978,
        month = jan,
       volume = {182},
        pages = {147-156},
          doi = {10.1093/mnras/182.2.147},
       adsurl = {https://ui.adsabs.harvard.edu/abs/1978MNRAS.182..147B}
}

@ARTICLE{Heesen14,
       author = {{Heesen}, Volker and {Brinks}, Elias and {Leroy}, Adam K. and {Heald}, George and {Braun}, Robert and {Bigiel}, Frank and {Beck}, Rainer},
        title = "{The Radio Continuum-Star Formation Rate Relation in WSRT SINGS Galaxies}",
      journal = {\aj},
         year = 2014,
        month = may,
       volume = {147},
       number = {5},
          eid = {103},
        pages = {103},
          doi = {10.1088/0004-6256/147/5/103},
archivePrefix = {arXiv},
       eprint = {1402.1711},
 primaryClass = {astro-ph.GA},
       adsurl = {https://ui.adsabs.harvard.edu/abs/2014AJ....147..103H}
}

@ARTICLE{Schleicher13,
       author = {{Schleicher}, Dominik R.~G. and {Beck}, Rainer},
        title = "{A new interpretation of the far-infrared - radio correlation and the expected breakdown at high redshift}",
      journal = {\aap},
         year = 2013,
        month = aug,
       volume = {556},
          eid = {A142},
        pages = {A142},
          doi = {10.1051/0004-6361/201321707},
archivePrefix = {arXiv},
       eprint = {1306.6652},
 primaryClass = {astro-ph.CO},
       adsurl = {https://ui.adsabs.harvard.edu/abs/2013A&A...556A.142S}
}

@ARTICLE{Koulouridis21,
       author = {{Koulouridis}, E. and {Clerc}, N. and {Sadibekova}, T. and {Chira}, M. and {Drigga}, E. and {Faccioli}, L. and {Le F{\`e}vre}, J.~P. and {Garrel}, C. and {Gaynullina}, E. and {Gkini}, A. and {Kosiba}, M. and {Pacaud}, F. and {Pierre}, M. and {Ridl}, J. and {Tazhenova}, K. and {Adami}, C. and {Altieri}, B. and {Baguley}, J.-C. and {Cabanac}, R. and {Cucchetti}, E. and {Khalikova}, A. and {Lieu}, M. and {Melin}, J.-B. and {Molham}, M. and {Ramos-Ceja}, M.~E. and {Soucail}, G. and {Takey}, A. and {Valtchanov}, I.},
        title = "{The X-CLASS survey: A catalogue of 1646 X-ray-selected galaxy clusters up to z {\ensuremath{\sim}} 1.5}",
      journal = {\aap},
         year = 2021,
        month = aug,
       volume = {652},
          eid = {A12},
        pages = {A12},
          doi = {10.1051/0004-6361/202140566},
archivePrefix = {arXiv},
       eprint = {2104.06617},
 primaryClass = {astro-ph.CO},
       adsurl = {https://ui.adsabs.harvard.edu/abs/2021A&A...652A..12K}
}

@ARTICLE{Vargas18,
       author = {{Vargas}, Carlos J. and {Mora-Partiarroyo}, Silvia Carolina and {Schmidt}, Philip and {Rand}, Richard J. and {Stein}, Yelena and {Walterbos}, Ren{\'e} A.~M. and {Wang}, Q. Daniel and {Basu}, Aritra and {Patterson}, Maria and {Kepley}, Amanda and {Beck}, Rainer and {Irwin}, Judith and {Heald}, George and {Li}, Jiangtao and {Wiegert}, Theresa},
        title = "{CHANG-ES X: Spatially Resolved Separation of Thermal Contribution from Radio Continuum Emission in Edge-on Galaxies}",
      journal = {\apj},
         year = 2018,
        month = feb,
       volume = {853},
       number = {2},
          eid = {128},
        pages = {128},
          doi = {10.3847/1538-4357/aaa47f},
archivePrefix = {arXiv},
       eprint = {1801.01892},
 primaryClass = {astro-ph.GA},
       adsurl = {https://ui.adsabs.harvard.edu/abs/2018ApJ...853..128V}
}

@ARTICLE{Irwin24b,
       author = {{Irwin}, Judith and {Cook}, Tanden and {Stein}, Michael and {Dettmar}, Ralf-Juergen and {Heesen}, Volker and {Wang}, Q. Daniel and {Wiegert}, Theresa and {Stein}, Yelena and {Vargas}, Carlos},
        title = "{CHANG-ES. XXXII. Spatially Resolved Thermal{\textendash}Nonthermal Separation from Radio Data Alone{\textemdash}New Probes into NGC 3044 and NGC 5775}",
      journal = {\aj},
         year = 2024,
        month = sep,
       volume = {168},
       number = {3},
          eid = {138},
        pages = {138},
          doi = {10.3847/1538-3881/ad660b},
archivePrefix = {arXiv},
       eprint = {2407.16442},
 primaryClass = {astro-ph.GA},
       adsurl = {https://ui.adsabs.harvard.edu/abs/2024AJ....168..138I}
}

@ARTICLE{N4666,
       author = {{Stein}, Y. and {Dettmar}, R. -J. and {Irwin}, J. and {Beck}, R. and {We{\.z}gowiec}, M. and {Miskolczi}, A. and {Krause}, M. and {Heesen}, V. and {Wiegert}, T. and {Heald}, G. and {Walterbos}, R.~A.~M. and {Li}, J. -T. and {Soida}, M.},
        title = "{CHANG-ES. XIII. Transport processes and the magnetic fields of NGC 4666: indication of a reversing disk magnetic field}",
      journal = {\aap},
         year = 2019,
        month = mar,
       volume = {623},
          eid = {A33},
        pages = {A33},
          doi = {10.1051/0004-6361/201834515},
archivePrefix = {arXiv},
       eprint = {1901.08090},
 primaryClass = {astro-ph.GA},
       adsurl = {https://ui.adsabs.harvard.edu/abs/2019A&A...623A..33S}
}

@ARTICLE{HImap,
       author = {{Heald}, G. and {J{\'o}zsa}, G. and {Serra}, P. and {Zschaechner}, L. and {Rand}, R. and {Fraternali}, F. and {Oosterloo}, T. and {Walterbos}, R. and {J{\"u}tte}, E. and {Gentile}, G.},
        title = "{The Westerbork Hydrogen Accretion in LOcal GAlaxieS (HALOGAS) survey. I. Survey description and pilot observations}",
      journal = {\aap},
         year = 2011,
        month = feb,
       volume = {526},
          eid = {A118},
        pages = {A118},
          doi = {10.1051/0004-6361/201015938},
archivePrefix = {arXiv},
       eprint = {1012.0816},
 primaryClass = {astro-ph.CO},
       adsurl = {https://ui.adsabs.harvard.edu/abs/2011A&A...526A.118H}
}

@ARTICLE{Condon92,
       author = {{Condon}, J.~J.},
        title = "{Radio emission from normal galaxies.}",
      journal = {\araa},
         year = 1992,
        month = jan,
       volume = {30},
        pages = {575-611},
          doi = {10.1146/annurev.aa.30.090192.003043},
       adsurl = {https://ui.adsabs.harvard.edu/abs/1992ARA&A..30..575C}
}

@ARTICLE{Bell03,
       author = {{Bell}, Eric F.},
        title = "{Estimating Star Formation Rates from Infrared and Radio Luminosities: The Origin of the Radio-Infrared Correlation}",
      journal = {\apj},
         year = 2003,
        month = apr,
       volume = {586},
       number = {2},
        pages = {794-813},
          doi = {10.1086/367829},
archivePrefix = {arXiv},
       eprint = {astro-ph/0212121},
 primaryClass = {astro-ph},
       adsurl = {https://ui.adsabs.harvard.edu/abs/2003ApJ...586..794B}
}

@ARTICLE{Murphy11,
       author = {{Murphy}, E.~J. and {Condon}, J.~J. and {Schinnerer}, E. and {Kennicutt}, R.~C. and {Calzetti}, D. and {Armus}, L. and {Helou}, G. and {Turner}, J.~L. and {Aniano}, G. and {Beir{\~a}o}, P. and {Bolatto}, A.~D. and {Brandl}, B.~R. and {Croxall}, K.~V. and {Dale}, D.~A. and {Donovan Meyer}, J.~L. and {Draine}, B.~T. and {Engelbracht}, C. and {Hunt}, L.~K. and {Hao}, C. -N. and {Koda}, J. and {Roussel}, H. and {Skibba}, R. and {Smith}, J. -D.~T.},
        title = "{Calibrating Extinction-free Star Formation Rate Diagnostics with 33 GHz Free-free Emission in NGC 6946}",
      journal = {\apj},
         year = 2011,
        month = aug,
       volume = {737},
       number = {2},
          eid = {67},
        pages = {67},
          doi = {10.1088/0004-637X/737/2/67},
archivePrefix = {arXiv},
       eprint = {1105.4877},
 primaryClass = {astro-ph.CO},
       adsurl = {https://ui.adsabs.harvard.edu/abs/2011ApJ...737...67M}
}

@ARTICLE{Heesen22,
       author = {{Heesen}, V. and {Staffehl}, M. and {Basu}, A. and {Beck}, R. and {Stein}, M. and {Tabatabaei}, F.~S. and {Hardcastle}, M.~J. and {Chy{\.z}y}, K.~T. and {Shimwell}, T.~W. and {Adebahr}, B. and {Beswick}, R. and {Bomans}, D.~J. and {Botteon}, A. and {Brinks}, E. and {Br{\"u}ggen}, M. and {Dettmar}, R. -J. and {Drabent}, A. and {de Gasperin}, F. and {G{\"u}rkan}, G. and {Heald}, G.~H. and {Horellou}, C. and {Nikiel-Wroczynski}, B. and {Paladino}, R. and {Piotrowska}, J. and {R{\"o}ttgering}, H.~J.~A. and {Smith}, D.~J.~B. and {Tasse}, C.},
        title = "{Nearby galaxies in the LOFAR Two-metre Sky Survey. I. Insights into the non-linearity of the radio-SFR relation}",
      journal = {\aap},
         year = 2022,
        month = aug,
       volume = {664},
          eid = {A83},
        pages = {A83},
          doi = {10.1051/0004-6361/202142878},
archivePrefix = {arXiv},
       eprint = {2204.00635},
 primaryClass = {astro-ph.GA},
       adsurl = {https://ui.adsabs.harvard.edu/abs/2022A&A...664A..83H}
}

@ARTICLE{Heesen24b,
       author = {{Heesen}, V. and {Schulz}, S. and {Br{\"u}ggen}, M. and {Edler}, H.~W. and {Stein}, M. and {Paladino}, R. and {Boselli}, A. and {Ignesti}, A. and {Fossati}, M. and {Dettmar}, R. -J.},
        title = "{Nearby galaxies in the LOFAR Two-metre Sky Survey. III. Influence of cosmic-ray transport on the radio-SFR relation}",
      journal = {\aap},
         year = 2024,
        month = feb,
       volume = {682},
          eid = {A83},
        pages = {A83},
          doi = {10.1051/0004-6361/202347394},
archivePrefix = {arXiv},
       eprint = {2309.05732},
 primaryClass = {astro-ph.GA},
       adsurl = {https://ui.adsabs.harvard.edu/abs/2024A&A...682A..83H}
}

@ARTICLE{Henriksen21,
       author = {{Henriksen}, R.~N. and {Irwin}, Judith},
        title = "{Turbulent Magnetic Dynamos with Halo Lags, Winds, and Jets}",
      journal = {\apj},
         year = 2021,
        month = oct,
       volume = {920},
       number = {2},
          eid = {133},
        pages = {133},
          doi = {10.3847/1538-4357/ac173f},
archivePrefix = {arXiv},
       eprint = {2107.10874},
 primaryClass = {astro-ph.GA},
       adsurl = {https://ui.adsabs.harvard.edu/abs/2021ApJ...920..133H}
}

@ARTICLE{Muller11,
       author = {{M{\"u}ller-S{\'a}nchez}, F. and {Prieto}, M.~A. and {Hicks}, E.~K.~S. and {Vives-Arias}, H. and {Davies}, R.~I. and {Malkan}, M. and {Tacconi}, L.~J. and {Genzel}, R.},
        title = "{Outflows from Active Galactic Nuclei: Kinematics of the Narrow-line and Coronal-line Regions in Seyfert Galaxies}",
      journal = {\apj},
         year = 2011,
        month = oct,
       volume = {739},
       number = {2},
          eid = {69},
        pages = {69},
          doi = {10.1088/0004-637X/739/2/69},
archivePrefix = {arXiv},
       eprint = {1107.3140},
 primaryClass = {astro-ph.CO},
       adsurl = {https://ui.adsabs.harvard.edu/abs/2011ApJ...739...69M}
}

@ARTICLE{LoTSS,
       author = {{Shimwell}, T.~W. and {R{\"o}ttgering}, H.~J.~A. and {Best}, P.~N. and {Williams}, W.~L. and {Dijkema}, T.~J. and {de Gasperin}, F. and {Hardcastle}, M.~J. and {Heald}, G.~H. and {Hoang}, D.~N. and {Horneffer}, A. and et al.},
        title = "{The LOFAR Two-metre Sky Survey. I. Survey description and preliminary data release}",
      journal = {\aap},
         year = 2017,
        month = feb,
       volume = {598},
          eid = {A104},
        pages = {A104},
          doi = {10.1051/0004-6361/201629313},
archivePrefix = {arXiv},
       eprint = {1611.02700},
 primaryClass = {astro-ph.IM},
       adsurl = {https://ui.adsabs.harvard.edu/abs/2017A&A...598A.104S}
}

@ARTICLE{Li08,
       author = {{Li}, Jiang-Tao and {Li}, Zhiyuan and {Wang}, Q. Daniel and {Irwin}, Judith A. and {Rossa}, Joern},
        title = "{Chandra observation of the edge-on spiral NGC 5775: probing the hot galactic disc/halo connection}",
      journal = {\mnras},
         year = 2008,
        month = oct,
       volume = {390},
       number = {1},
        pages = {59-70},
          doi = {10.1111/j.1365-2966.2008.13749.x},
archivePrefix = {arXiv},
       eprint = {0807.3587},
 primaryClass = {astro-ph},
       adsurl = {https://ui.adsabs.harvard.edu/abs/2008MNRAS.390...59L}
}

@ARTICLE{Li13b,
       author = {{Li}, Jiang-Tao and {Wang}, Q. Daniel},
        title = "{Chandra survey of nearby highly inclined disc galaxies - II. Correlation analysis of galactic coronal properties}",
      journal = {\mnras},
         year = 2013,
        month = nov,
       volume = {435},
       number = {4},
        pages = {3071-3084},
          doi = {10.1093/mnras/stt1501},
archivePrefix = {arXiv},
       eprint = {1308.1933},
 primaryClass = {astro-ph.CO},
       adsurl = {https://ui.adsabs.harvard.edu/abs/2013MNRAS.435.3071L}
}

@ARTICLE{Li16,
       author = {{Li}, Jiang-Tao and {Beck}, Rainer and {Dettmar}, Ralf-J{\"u}rgen and {Heald}, George and {Irwin}, Judith and {Johnson}, Megan and {Kepley}, Amanda A. and {Krause}, Marita and {Murphy}, E.~J. and {Orlando}, Elena and {Rand}, Richard J. and {Strong}, A.~W. and {Vargas}, Carlos J. and {Walterbos}, Rene and {Wang}, Q. Daniel and {Wiegert}, Theresa},
        title = "{CHANG-ES - VI. Probing Supernova energy deposition in spiral galaxies through multiwavelength relationships}",
      journal = {\mnras},
         year = 2016,
        month = feb,
       volume = {456},
       number = {2},
        pages = {1723-1738},
          doi = {10.1093/mnras/stv2757},
archivePrefix = {arXiv},
       eprint = {1511.06673},
 primaryClass = {astro-ph.GA},
       adsurl = {https://ui.adsabs.harvard.edu/abs/2016MNRAS.456.1723L}
}

@ARTICLE{Li22,
       author = {{Li}, Jiang-Tao and {Wang}, Q. Daniel and {Wiegert}, Theresa and {Bregman}, Joel N. and {Beck}, Rainer and {Damas-Segovia}, Ancor and {Irwin}, Judith A. and {Ji}, Li and {Stein}, Yelena and {Sun}, Wei and {Yang}, Yang},
        title = "{CHANG-ES XXIX: the sub-kpc nuclear bubble of NGC 4438}",
      journal = {\mnras},
         year = 2022,
        month = sep,
       volume = {515},
       number = {2},
        pages = {2483-2495},
          doi = {10.1093/mnras/stac837},
archivePrefix = {arXiv},
       eprint = {2205.12343},
 primaryClass = {astro-ph.HE},
       adsurl = {https://ui.adsabs.harvard.edu/abs/2022MNRAS.515.2483L}
}

@ARTICLE{Li24a,
       author = {{Li}, Jiang-Tao and {Sun}, Wei and {Ji}, Li and {Yang}, Yang},
        title = "{Pressure Balance and Energy Budget of the Nuclear Superbubble of NGC 3079}",
      journal = {\apj},
         year = 2024,
        month = may,
       volume = {966},
       number = {2},
          eid = {239},
        pages = {239},
          doi = {10.3847/1538-4357/ad3af2},
archivePrefix = {arXiv},
       eprint = {2404.03879},
 primaryClass = {astro-ph.GA},
       adsurl = {https://ui.adsabs.harvard.edu/abs/2024ApJ...966..239L}
}

@ARTICLE{Li24b,
       author = {{Li}, Jiang-Tao and {Lu}, Li-Yuan and {Qu}, Zhijie and {Benjamin}, Robert A. and {Bregman}, Joel N. and {Dettmar}, Ralf-J{\"u}rgen and {English}, Jayanne and {Fang}, Taotao and {Irwin}, Judith A. and {Jiang}, Yan and {Li}, Hui and {Liu}, Guilin and {Martini}, Paul and {Rand}, Richard J. and {Stein}, Yelena and {Strong}, Andrew W. and {Vargas}, Carlos J. and {Wang}, Q. Daniel and {Wang}, Jing and {Wiegert}, Theresa and {Xu}, Jianghui and {Yang}, Yang},
        title = "{eDIG-CHANGES. II. Project Design and Initial Results on NGC 3556}",
      journal = {\apj},
         year = 2024,
        month = jun,
       volume = {967},
       number = {2},
          eid = {78},
        pages = {78},
          doi = {10.3847/1538-4357/ad3cd8},
archivePrefix = {arXiv},
       eprint = {2404.05628},
 primaryClass = {astro-ph.GA},
       adsurl = {https://ui.adsabs.harvard.edu/abs/2024ApJ...967...78L}
}

@ARTICLE{Lu23,
       author = {{Lu}, Li-Yuan and {Li}, Jiang-Tao and {Vargas}, Carlos J. and {Beck}, Rainer and {Bregman}, Joel N. and {Dettmar}, Ralf-J{\"u}rgen and {English}, Jayanne and {Fang}, Taotao and {Heald}, George H. and {Li}, Hui and {Qu}, Zhijie and {Rand}, Richard J. and {Stein}, Michael and {Wang}, Q. Daniel and {Wang}, Jing and {Wiegert}, Theresa and {Zheng}, Yun},
        title = "{eDIG-CHANGES I: extended H{\ensuremath{\alpha}} emission from the extraplanar diffuse ionized gas (eDIG) around CHANG-ES galaxies}",
      journal = {\mnras},
         year = 2023,
        month = mar,
       volume = {519},
       number = {4},
        pages = {6098-6110},
          doi = {10.1093/mnras/stad006},
archivePrefix = {arXiv},
       eprint = {2212.14824},
 primaryClass = {astro-ph.GA},
       adsurl = {https://ui.adsabs.harvard.edu/abs/2023MNRAS.519.6098L}
}

@ARTICLE{Yang24,
       author = {{Yang}, Yang and {Li}, Jiang-Tao and {Wiegert}, Theresa and {Li}, Zhiyuan and {Guo}, Fulai and {Irwin}, Judith and {Wang}, Q. Daniel and {Dettmar}, Ralf-Juergen and {Beck}, Rainer and {English}, Jayanne and {Ji}, Li},
        title = "{CHANG-ES. XXX. 10 kpc Radio Lobes in the Sombrero Galaxy}",
      journal = {\apj},
         year = 2024,
        month = may,
       volume = {966},
       number = {2},
          eid = {213},
        pages = {213},
          doi = {10.3847/1538-4357/ad37fd},
archivePrefix = {arXiv},
       eprint = {2403.16682},
 primaryClass = {astro-ph.GA},
       adsurl = {https://ui.adsabs.harvard.edu/abs/2024ApJ...966..213Y}
}

@ARTICLE{Irwin19b,
       author = {{Irwin}, Judith and {Damas-Segovia}, Ancor and {Krause}, Marita and {Miskolczi}, Arpad and {Li}, Jiangtao and {Stein}, Yelena and {English}, Jayanne and {Henriksen}, Richard and {Beck}, Rainer and {Wiegert}, Theresa and {Dettmar}, Ralf-J{\"u}rgen},
        title = "{CHANG-ES: XVIII{\textemdash}The CHANG-ES Survey and Selected Results}",
      journal = {Galaxies},
         year = 2019,
        month = mar,
       volume = {7},
       number = {1},
          eid = {42},
        pages = {42},
          doi = {10.3390/galaxies7010042},
archivePrefix = {arXiv},
       eprint = {1903.11042},
 primaryClass = {astro-ph.GA},
       adsurl = {https://ui.adsabs.harvard.edu/abs/2019Galax...7...42I}
}

@ARTICLE{Irwin24,
       author = {{Irwin}, Judith and {Beck}, Rainer and {Cook}, Tanden and {Dettmar}, Ralf-J{\"u}rgen and {English}, Jayanne and {Heesen}, Volker and {Henriksen}, Richard and {Jiang}, Yan and {Li}, Jiang-Tao and {Lu}, Li-Yuan and {Mele}, Crystal and {M{\"u}ller}, Ancla and {Murphy}, Eric and {Porter}, Troy and {Rand}, Richard and {Skeggs}, Nathan and {Stein}, Michael and {Stein}, Yelena and {Stil}, Jeroen and {Strong}, Andrew and {Walterbos}, Rene and {Wang}, Q. Daniel and {Wiegert}, Theresa and {Yang}, Yang},
        title = "{CHANG-ES XXXI{\textemdash}A Decade of CHANG-ES: What We Have Learned from Radio Observations of Edge-on Galaxies}",
      journal = {Galaxies},
         year = 2024,
        month = may,
       volume = {12},
       number = {3},
          eid = {22},
        pages = {22},
          doi = {10.3390/galaxies12030022},
       adsurl = {https://ui.adsabs.harvard.edu/abs/2024Galax..12...22I}
}

@ARTICLE{Zheng22,
       author = {{Zheng}, Yun and {Wang}, Jing and {Irwin}, Judith and {English}, Jayanne and {Ma}, Qingchuan and {Wang}, Ran and {Wang}, Ke and {Wang}, Q. Daniel and {Krause}, Marita and {Randriamampandry}, Toky H. and {Li}, Jiangtao and {Beck}, Rainer},
        title = "{CHANG-ES XXV: H I imaging of nearby edge-on galaxies - Data Release 4}",
      journal = {\mnras},
         year = 2022,
        month = jun,
       volume = {513},
       number = {1},
        pages = {1329-1353},
          doi = {10.1093/mnras/stac760},
archivePrefix = {arXiv},
       eprint = {2203.07818},
 primaryClass = {astro-ph.GA},
       adsurl = {https://ui.adsabs.harvard.edu/abs/2022MNRAS.513.1329Z}
}

@ARTICLE{Seta19,
       author = {{Seta}, Amit and {Beck}, Rainer},
        title = "{Revisiting the Equipartition Assumption in Star-Forming Galaxies}",
      journal = {Galaxies},
         year = 2019,
        month = apr,
       volume = {7},
       number = {2},
          eid = {45},
        pages = {45},
          doi = {10.3390/galaxies7020045},
archivePrefix = {arXiv},
       eprint = {1903.11856},
 primaryClass = {astro-ph.GA},
       adsurl = {https://ui.adsabs.harvard.edu/abs/2019Galax...7...45S}
}

@ARTICLE{Strong07,
       author = {{Strong}, Andrew W. and {Moskalenko}, Igor V. and {Ptuskin}, Vladimir S.},
        title = "{Cosmic-Ray Propagation and Interactions in the Galaxy}",
      journal = {Annual Review of Nuclear and Particle Science},
         year = 2007,
        month = nov,
       volume = {57},
       number = {1},
        pages = {285-327},
          doi = {10.1146/annurev.nucl.57.090506.123011},
archivePrefix = {arXiv},
       eprint = {astro-ph/0701517},
 primaryClass = {astro-ph},
       adsurl = {https://ui.adsabs.harvard.edu/abs/2007ARNPS..57..285S}
}

@article{Smith2001,
       author = {{Smith}, Randall K. and {Brickhouse}, Nancy S. and {Liedahl}, Duane A. and {Raymond}, John C.},
        title = "{Collisional Plasma Models with APEC/APED: Emission-Line Diagnostics of Hydrogen-like and Helium-like Ions}",
      journal = {\apjl},
         year = 2001,
        month = aug,
       volume = {556},
       number = {2},
        pages = {L91-L95},
          doi = {10.1086/322992},
archivePrefix = {arXiv},
       eprint = {astro-ph/0106478},
 primaryClass = {astro-ph},
       adsurl = {https://ui.adsabs.harvard.edu/abs/2001ApJ...556L..91S}
}

@ARTICLE{Luan2025,
       author = {{Luan}, Luan and {Wang}, Q. Daniel},
        title = "{Diffuse X-Ray Emission in M51: A Hierarchical Bayesian Spatially Resolved Spectral Analysis}",
      journal = {\apj},
         year = 2025,
        month = jun,
       volume = {986},
       number = {2},
          eid = {167},
        pages = {167},
          doi = {10.3847/1538-4357/adcf16},
archivePrefix = {arXiv},
       eprint = {2504.15641},
 primaryClass = {astro-ph.HE},
       adsurl = {https://ui.adsabs.harvard.edu/abs/2025ApJ...986..167L}
}

@article{HI4PI2016,
  author = {{HI4PI Collaboration} and N. Ben Bekhti and L. Fl{\"o}er and others},
  title = {HI4PI: A full-sky H\,{\sc i} survey based on EBHIS and GASS},
  journal = {A\&A},
  volume = {594},
  pages = {A116},
  year = {2016}
}

@ARTICLE{Suzuki2021,
       author = {{Suzuki}, H. and {Plucinsky}, P.~P. and {Gaetz}, T.~J. and {Bamba}, A.},
        title = "{Spatial and temporal variations of the Chandra ACIS particle-induced background and development of a spectral-model generation tool}",
      journal = {\aap},
         year = 2021,
        month = nov,
       volume = {655},
          eid = {A116},
        pages = {A116},
          doi = {10.1051/0004-6361/202141458},
archivePrefix = {arXiv},
       eprint = {2108.11234},
 primaryClass = {astro-ph.HE},
       adsurl = {https://ui.adsabs.harvard.edu/abs/2021A&A...655A.116S}
}

@INPROCEEDINGS{XSPEC1996,
       author = {{Arnaud}, K.~A.},
        title = "{XSPEC: The First Ten Years}",
    booktitle = {Astronomical Data Analysis Software and Systems V},
         year = 1996,
       editor = {{Jacoby}, George H. and {Barnes}, Jeannette},
       series = {Astronomical Society of the Pacific Conference Series},
       volume = {101},
        month = jan,
        pages = {17},
       adsurl = {https://ui.adsabs.harvard.edu/abs/1996ASPC..101...17A}
}

@ARTICLE{Chiu25,
       author = {{Chiu}, H. -H. Sandy and {Ruszkowski}, Mateusz and {Werhahn}, Maria and {Pfrommer}, Christoph and {Thomas}, Timon},
        title = "{Robust magnetic field estimates in star-forming galaxies with the equipartition formula in the absence of equipartition}",
      journal = {arXiv e-prints},
         year = 2025,
        month = oct,
          eid = {arXiv:2510.03229},
        pages = {arXiv:2510.03229},
          doi = {10.48550/arXiv.2510.03229},
archivePrefix = {arXiv},
       eprint = {2510.03229},
 primaryClass = {astro-ph.GA},
       adsurl = {https://ui.adsabs.harvard.edu/abs/2025arXiv251003229C}
}

@ARTICLE{Yang22,
       author = {{Yang}, Yang and {Irwin}, Judith and {Li}, Jiangtao and {Wiegert}, Theresa and {Wang}, Q. Daniel and {Sun}, Wei and {Damas-Segovia}, A. and {Li}, Zhiyuan and {Shen}, Zhiqiang and {Walterbos}, Ren{\'e} A.~M. and {Vargas}, Carlos J.},
        title = "{CHANG-ES. XXIV. First Detection of a Radio Nuclear Ring and Potential LLAGN in NGC 5792}",
      journal = {\apj},
         year = 2022,
        month = mar,
       volume = {927},
       number = {1},
          eid = {4},
        pages = {4},
          doi = {10.3847/1538-4357/ac4ae7},
archivePrefix = {arXiv},
       eprint = {2201.05384},
 primaryClass = {astro-ph.GA},
       adsurl = {https://ui.adsabs.harvard.edu/abs/2022ApJ...927....4Y}
}

@ARTICLE{Mart23,
       author = {{Mart{\'\i}nez-Lombilla}, Cristina and {Infante-Sainz}, Ra{\'u}l and {Jim{\'e}nez-Ibarra}, Felipe and {Knapen}, Johan H. and {Trujillo}, Ignacio and {Comer{\'o}n}, S{\'e}bastien and {Borlaff}, Alejandro S. and {Rom{\'a}n}, Javier},
        title = "{The truncation of the disk of NGC 4565. Detected up to z = 4 kpc, with star formation, and affected by the warp}",
      journal = {\aap},
         year = 2023,
        month = oct,
       volume = {678},
          eid = {A62},
        pages = {A62},
          doi = {10.1051/0004-6361/202346280},
archivePrefix = {arXiv},
       eprint = {2307.01106},
 primaryClass = {astro-ph.GA},
       adsurl = {https://ui.adsabs.harvard.edu/abs/2023A&A...678A..62M}
}

@ARTICLE{NGC1097,
       author = {{Beck}, R. and {Fletcher}, A. and {Shukurov}, A. and {Snodin}, A. and {Sokoloff}, D.~D. and {Ehle}, M. and {Moss}, D. and {Shoutenkov}, V.},
        title = "{Magnetic fields in barred galaxies. IV. NGC 1097 and NGC 1365}",
      journal = {\aap},
         year = 2005,
        month = dec,
       volume = {444},
       number = {3},
        pages = {739-765},
          doi = {10.1051/0004-6361:20053556},
archivePrefix = {arXiv},
       eprint = {astro-ph/0508485},
 primaryClass = {astro-ph},
       adsurl = {https://ui.adsabs.harvard.edu/abs/2005A&A...444..739B}
}

@ARTICLE{Gemini,
       author = {{Gemini Team} and {Anil}, Rohan and {Borgeaud}, Sebastian and {Alayrac}, Jean-Baptiste and {Yu}, Jiahui and {Soricut}, Radu and {Schalkwyk}, Johan and {Dai}, Andrew M. and {Hauth}, Anja and {Millican}, Katie and {Silver}, David and {Johnson}, Melvin and {Antonoglou}, Ioannis and {Schrittwieser}, Julian and {Glaese}, Amelia and {Chen}, Jilin and {Pitler}, Emily and {Lillicrap}, Timothy and {Lazaridou}, Angeliki and {Firat}, Orhan and {Molloy}, James and {Isard}, Michael and {Barham}, Paul R. and {Hennigan}, Tom and {Lee}, Benjamin and {Viola}, Fabio and {Reynolds}, Malcolm and {Xu}, Yuanzhong and {Doherty}, Ryan and {Collins}, Eli and {Meyer}, Clemens and {Rutherford}, Eliza and {Moreira}, Erica and {Ayoub}, Kareem and {Goel}, Megha and {Krawczyk}, Jack and {Du}, Cosmo and {Chi}, Ed and {Cheng}, Heng-Tze and {Ni}, Eric and {Shah}, Purvi and {Kane}, Patrick and {Chan}, Betty and {Faruqui}, Manaal and {Severyn}, Aliaksei and {Lin}, Hanzhao and {Li}, YaGuang and {Cheng}, Yong and {Ittycheriah}, Abe and {Mahdieh}, Mahdis and {Chen}, Mia and {Sun}, Pei and {Tran}, Dustin and {Bagri}, Sumit and {Lakshminarayanan}, Balaji and {Liu}, Jeremiah and {Orban}, Andras and {G{\"u}ra}, Fabian and {Zhou}, Hao and {Song}, Xinying and {Boffy}, Aurelien and {Ganapathy}, Harish and {Zheng}, Steven and {Choe}, HyunJeong and {Weisz}, {\'A}goston and {Zhu}, Tao and {Lu}, Yifeng and {Gopal}, Siddharth and {Kahn}, Jarrod and {Kula}, Maciej and {Pitman}, Jeff and {Shah}, Rushin and {Taropa}, Emanuel and {Al Merey}, Majd and {Baeuml}, Martin and {Chen}, Zhifeng and {El Shafey}, Laurent and {Zhang}, Yujing and {Sercinoglu}, Olcan and {Tucker}, George and {Piqueras}, Enrique and {Krikun}, Maxim and {Barr}, Iain and {Savinov}, Nikolay and {Danihelka}, Ivo and {Roelofs}, Becca and {White}, Ana{\"\i}s and {Andreassen}, Anders and {von Glehn}, Tamara and {Yagati}, Lakshman and {Kazemi}, Mehran and {Gonzalez}, Lucas and {Khalman}, Misha and {Sygnowski}, Jakub and {Frechette}, Alexandre and {Smith}, Charlotte and {Culp}, Laura and {Proleev}, Lev and {Luan}, Yi and {Chen}, Xi and {Lottes}, James and {Schucher}, Nathan and {Lebron}, Federico and {Rrustemi}, Alban and {Clay}, Natalie and {Crone}, Phil and {Kocisky}, Tomas and {Zhao}, Jeffrey and {Perz}, Bartek and {Yu}, Dian and {Howard}, Heidi and {Bloniarz}, Adam and {Rae}, Jack W. and {Lu}, Han and {Sifre}, Laurent and {Maggioni}, Marcello and {Alcober}, Fred and {Garrette}, Dan and {Barnes}, Megan and {Thakoor}, Shantanu and {Austin}, Jacob and {Barth-Maron}, Gabriel and {Wong}, William and {Joshi}, Rishabh and {Chaabouni}, Rahma and {Fatiha}, Deeni and {Ahuja}, Arun and {Singh Tomar}, Gaurav and {Senter}, Evan and {Chadwick}, Martin and {Kornakov}, Ilya and {Attaluri}, Nithya and {Iturrate}, I{\~n}aki and {Liu}, Ruibo and {Li}, Yunxuan and {Cogan}, Sarah and {Chen}, Jeremy and {Jia}, Chao and {Gu}, Chenjie and {Zhang}, Qiao and {Grimstad}, Jordan and {Jakse Hartman}, Ale and {Garcia}, Xavier and {Sankaranarayana Pillai}, Thanumalayan and {Devlin}, Jacob and {Laskin}, Michael and {de Las Casas}, Diego and {Valter}, Dasha and {Tao}, Connie and {Blanco}, Lorenzo and {Puigdom{\`e}nech Badia}, Adri{\`a} and {Reitter}, David and {Chen}, Mianna and {Brennan}, Jenny and {Rivera}, Clara and {Brin}, Sergey and {Iqbal}, Shariq and {Surita}, Gabriela and {Labanowski}, Jane and {Rao}, Abhi and {Winkler}, Stephanie and {Parisotto}, Emilio and {Gu}, Yiming and {Olszewska}, Kate and {Addanki}, Ravi and {Miech}, Antoine and {Louis}, Annie and {Teplyashin}, Denis and {Brown}, Geoff and {Catt}, Elliot and {Balaguer}, Jan and {Xiang}, Jackie and {Wang}, Pidong and {Ashwood}, Zoe and {Briukhov}, Anton and {Webson}, Albert and {Ganapathy}, Sanjay and {Sanghavi}, Smit and {Kannan}, Ajay and {Chang}, Ming-Wei and {Stjerngren}, Axel and {Djolonga}, Josip and {Sun}, Yuting and {Bapna}, Ankur and {Aitchison}, Matthew and {Pejman}, Pedram and {Michalewski}, Henryk and {Yu}, Tianhe and {Wang}, Cindy and {Love}, Juliette and {Ahn}, Junwhan and {Bloxwich}, Dawn and {Han}, Kehang and {Humphreys}, Peter and {Sellam}, Thibault and {Bradbury}, James and {Godbole}, Varun and {Samangooei}, Sina and {Damoc}, Bogdan and {Kaskasoli}, Alex},
        title = "{Gemini: A Family of Highly Capable Multimodal Models}",
      journal = {arXiv e-prints},
         year = 2023,
        month = dec,
          eid = {arXiv:2312.11805},
        pages = {arXiv:2312.11805},
          doi = {10.48550/arXiv.2312.11805},
archivePrefix = {arXiv},
       eprint = {2312.11805},
 primaryClass = {cs.CL},
       adsurl = {https://ui.adsabs.harvard.edu/abs/2023arXiv231211805G}
}

@ARTICLE{LoTSS_DR3,
       author = {{Shimwell}, T.~W. and {Hardcastle}, M.~J. and {Tasse}, C. and {Drabent}, A. and {Botteon}, A. and {Williams}, W.~L. and {Best}, P.~N. and {R{\"o}ttgering}, H.~J.~A. and {Br{\"u}ggen}, M. and {Brunetti}, G. and {Callingham}, J.~R. and {Chy{\.z}y}, K.~T. and {Conway}, J.~E. and {De Gasperin}, F. and {Haverkorn}, M. and {Horellou}, C. and {Jackson}, N. and {Miley}, G.~K. and {Morabito}, L.~K. and {Morganti}, R. and {O'Sullivan}, S.~P. and {Schwarz}, D.~J. and {Smith}, D.~J.~B. and {van Weeren}, R.~J. and {Vedantham}, H.~K. and {White}, G.~J. and {Ahmadi}, A. and {Alegre}, L. and {Arias}, M. and {Asabere}, B. and {Bahr-Kalus}, B. and {Barkus}, B. and {Bilicki}, M. and {B{\"o}hme}, L. and {Brentjens}, M. and {Brienza}, M. and {Bomans}, D.~J. and {Bonafede}, A. and {Bonato}, M. and {Bonnassieux}, E. and {Boxelaar}, J.~M. and {Camera}, S. and {Cassano}, R. and {Chilufya}, J. and {Cianfaglione}, M. and {Croston}, J.~H. and {Cuciti}, V. and {Dabhade}, P. and {De Rubeis}, E. and {de Jong}, J.~M.~G.~H.~J. and {Dallacasa}, D. and {Dettmar}, R.~J. and {Duncan}, K.~J. and {Di Gennaro}, G. and {Edler}, H.~W. and {Groeneveld}, C. and {G{\"u}rkan}, G. and {Hajduk}, M. and {Hale}, C.~L. and {Heesen}, V. and {Hoang}, D.~N. and {Hoeft}, M. and {Holties}, H. and {Horton}, M.~A. and {Iacobelli}, M. and {Jamrozy}, M. and {Jarvis}, M.~J. and {Jelic}, V. and {Kadler}, M. and {Kondapally}, R. and {Kunert-Bajraszewska}, M. and {Loose}, M. and {Magliocchetti}, M. and {Ma{\l}ek}, K. and {Manzano}, C. and {McKean}, J.~P. and {Mevius}, M. and {Mingo}, B. and {Miskolczi}, A. and {Misra}, A. and {Mold{\'o}n}, J. and {Nair}, D.~G. and {Nakoneczny}, S.~J. and {Orru}, E. and {Pashapour-Ahmadabadi}, M. and {Pasini}, T. and {Petley}, J. and {Pierce}, J.~C.~S. and {Prandoni}, I. and {Rafferty}, D. and {Rajpurohit}, K. and {Riseley}, C.~J. and {Roberts}, I.~D. and {Sethi}, S. and {Shulevski}, A. and {Stein}, M. and {Stuardi}, C. and {Sweijen}, F. and {ter Veen}, S. and {Timmerman}, R. and {Vaccari}, M. and {Wijnholds}, S.},
        title = "{The LOFAR Two-metre Sky Survey: VII. Third Data Release}",
      journal = {arXiv e-prints},
         year = 2026,
        month = feb,
          eid = {arXiv:2602.15949},
        pages = {arXiv:2602.15949},
archivePrefix = {arXiv},
       eprint = {2602.15949},
 primaryClass = {astro-ph.GA},
       adsurl = {https://ui.adsabs.harvard.edu/abs/2026arXiv260215949S}
}

@ARTICLE{Pol_bias,
       author = {{George}, Samuel J. and {Stil}, Jeroen M. and {Keller}, Ben W.},
        title = "{Detection Thresholds and Bias Correction in Polarized Intensity}",
      journal = {\pasa},
         year = 2012,
        month = oct,
       volume = {29},
       number = {3},
        pages = {214-220},
          doi = {10.1071/AS11027},
archivePrefix = {arXiv},
       eprint = {1106.5362},
 primaryClass = {astro-ph.IM},
       adsurl = {https://ui.adsabs.harvard.edu/abs/2012PASA...29..214G}
}
\bibliographystyle{aasjournalv7}

\end{document}